# An analysis of degradation in low-cost particulate matter sensors

Priyanka deSouza[1,2*], Karoline Barkjohn[3], Andrea Clements[3], Jenny Lee[4], Ralph Kahn[5], Ben Crawford[6], Patrick Kinney[7]

1: Department of Urban and Regional Planning, University of Colorado Denver, Denver CO, 80202, USA
2: CU Population Center, University of Colorado Boulder, Boulder CO, 80302, USA
3: Office of Research and Development, US Environmental Protection Agency, 109 T.W. Alexander Drive, Research Triangle Park, NC 27711, USA
4: Department of Biostatistics, Harvard T.H. Chan School of Public Health, Boston, MA 02115, USA
5: NASA Goddard Space Flight Center, Greenbelt, MA, 20771, USA
6: Department of Geography and Environmental Sciences, University of Colorado Denver, 80202, USA
7: Boston University School of Public Health, Boston, MA, 02118 USA

*: Corresponding author priyanka.desouza@ucdenver.edu

# Abstract

Low-cost sensors (LCS) are increasingly being used to measure fine particulate matter ($PM_{2.5}$) concentrations in cities around the world. One of the most commonly deployed LCS is the PurpleAir with ~ 15,000 sensors deployed in the United States. PurpleAir measurements are widely used by the public to evaluate $PM_{2.5}$ levels in their neighborhoods. PurpleAir measurements are also increasingly being integrated into models by researchers to develop large-scale estimates of $PM_{2.5}$. However, the change in sensor performance over time has not been well studied. It is important to understand the lifespan of these sensors to determine when they should be replaced, and when measurements from these devices should or should not be used for various applications. This paper fills in this gap by leveraging the fact that: (1) Each PurpleAir sensor is comprised of two identical sensors and the divergence between their measurements can be observed, and (2) There are numerous PurpleAir sensors within ~ 50 meters of regulatory monitors allowing for the comparison of measurements between these instruments. We propose empirically-derived degradation outcomes for the PurpleAir sensors and evaluate how these outcomes change over time. On average, we find that the number of 'flagged' measurements, where the two sensors within each PurpleAir sensor disagree, increases with time to ~ 4% after 4 years of operation. Approximately, 2 percent of all PurpleAir sensors were permanently degraded. The largest fraction of permanently degraded PurpleAir sensors appeared to be in the hot and humid climate zone, suggesting that sensors in this location may need to be replaced sooner. We also find that the bias of PurpleAir sensors, or the difference between corrected $PM_{2.5}$ levels and the corresponding reference measurements, changed over time by -0.12 µg/m$^3$ (95% CI: -0.13 µg/m$^3$, -0.11 µg/m$^3$) per year. The average bias increases dramatically after 3.5 years. Climate zone is a significant modifier of the association between degradation outcomes and time.

# Introduction

Poor air quality is currently the single largest environmental risk factor to human health in the world, with ambient air pollution responsible for 6.7 million premature deaths every year [1]. Accurate air quality data is crucial for tracking long-term trends in air quality levels, and for the development of effective pollution management plans. Levels of fine particulate matter ($PM_{2.5}$), a criteria pollutant that poses more danger to human health than other widespread pollutants [2], can vary over distances as small as ~ 10's of meters in complex urban environments [3,4]. Therefore, dense monitoring networks are often needed to capture relevant spatial variations. U.S EPA air quality monitoring networks use approved Federal Reference of Equivalent Method (FRM/FEM) monitors, the gold standard for measuring air pollutants. However these monitors are sparsely positioned across the US [5,6].

Low-cost sensors (LCS) (< $2,500 USD as defined by the U.S. EPA[7]) have the potential to capture concentrations of particulate matter (PM) in previously unmonitored locations and democratize air pollution information [8–14]. Measurements from these devices are increasingly being integrated into models to develop large-scale exposure assessments [15–17].

Most low-cost PM sensors rely on optical measurement techniques that introduce potential differences in mass estimations compared to reference monitors (i.e FRM/FEM monitors) [18–20]. Optical sensor methods do not directly measure mass concentrations; rather, they measure light scattering of particles having diameters typically > ~ 0.3 µm. Several assumptions are typically made to convert light scattering into mass concentrations that can introduce error into the results. In addition, data from LCS are impacted by environmental variables such as relative humidity (RH), temperature (T), and dewpoint (D). Many research groups have developed different correction techniques to correct the raw LCS measurements for PM sensors. Such models typically adjust for 1) systematic error, and 2) the dependencies of low-cost PM sensors measurements on RH, T and D.

Little work has been done to evaluate the performance of low-cost PM sensors over time. There is evidence that the performance of these instruments can be affected by high PM events which can impact later measurements if the sensors are not cleaned properly [21]. Although there has been some research evaluating drift in measurements from low-cost electrochemical gas sensors [22,23], there has been less work evaluating drift and degradation in low-cost PM sensors, and identifying which factors affect these outcomes. An understanding of degradation could lead to better protocols for correcting low-cost PM sensors and could provide users with information on when to service or replace their sensors or whether data should or should not be used for certain applications.

This paper evaluates the performance of the PurpleAir sensor, one of the most common low-cost PM sensors over time. We choose to conduct this analysis with PurpleAir because:
1) There is a sizable number of PurpleAir sensors within 50 meters of regulatory monitors that allows for comparison between PurpleAir measurements and reference data over time, and
2) Each PurpleAir sensor consists of two identical PM sensors making it possible to evaluate how the two sensors disagree over time, and the different factors that contribute to this disagreement.
3) Several studies have evaluated the short-term performance of the PurpleAir sensors in many different locations, under a variety of conditions around the world [24,25]. However, none of these studies have evaluated the performance of the PurpleAir sensors over time. We aim to fill in this gap.

# 2 Data and Methods

## 2.1 PurpleAir measurements

There are two main types of PurpleAir sensors available for purchase: PA-I and PA-II. PA-I sensors have one PM sensor component (Plantower PMS 1003) for PM measurement. Whereas, the PA-II PurpleAir sensor has two identical PM sensor components (Plantower PMS 5003 sensors) referred to as "Channel A" and "Channel B". In this study, measurements were restricted to PA-II PurpleAir sensors in order to compare Channels A and B. PA-II-Flex (which uses Plantower PMS 6003 PM sensors) were not used in this study as they were not made available until early 2022, after the dataset for this project was downloaded.

The PA-II PurpleAir sensor operates for 10 s at alternating intervals and provides 2-min averaged data (prior to 30 May 2019, this was 80 s averaged data). The Plantower sensor components measure $90^0$ light scattering with a laser at $680 \pm 10$ nm wavelength [26,27] and are factory calibrated using ambient aerosol across several cities in China [20]. The Plantower sensor reports estimated mass concentrations of particles with aerodynamic diameters < 1 μm ($PM_1$), < 2.5 μm ($PM_{2.5}$), and < 10 μm ($PM_{10}$). For each PM size fraction, the values are reported in two ways, labeled cf_1 and cf_atm, in the PurpleAir dataset, which match the "raw" Plantower outputs.

The ratio of cf_atm and cf_1 (i.e. [cf_atm]/ [cf_1]) is equal to 1 below $PM_{2.5}$ concentrations around 25 μg/m³ (as reported by the sensor) and then transition to a two-thirds ratio at a higher PM concentration (cf_1 concentrations are higher). The cf_atm data, displayed on the PurpleAir map, are the lower measurement of $PM_{2.5}$ and will be referred to as the "raw" data in this paper when making comparisons between initial and corrected datasets [26]. When a PurpleAir sensor is connected to the internet, data are sent to PurpleAir's data repository. Users can choose to make their data publicly viewable (public) or control data sharing (private). All PurpleAir sensors also report RH and T levels.

For this study, data from 14,927 PurpleAir sensors operating in the United States (excluding US territories) between 1 January, 2017 to 20 July 2021 were downloaded from the API at 15 minute time resolution. Note that a small number of PurpleAir sensors have been operational before 2017. However, given that the number of PurpleAir sensors increased dramatically from 2017 onwards, we choose Jan 1, 2017 as the start date of our analysis. Overall, 26.2% of dates had missing measurements, likely due to power outages or loss of WiFi that prevented the PurpleAir sensors from transmitting data. Of the sensors in our dataset, 2,989 were missing channel B data, leaving us with 483,511,216 measurements from 11,938 sensors with both channel A and channel B data. We removed all records with missing $PM_{2.5}$ measurements in cf_1 channels A and B (~0.9% of the data). We then removed all records with missing T and RH data (~ 2.6% of all data). Of the non-missing records, all measurements where $PM_{2.5}$ in cf_1 channels A and B were both > 1500 μg/m³ were removed as they correspond to conditions beyond the operating range of the PurpleAir sensor[18]. We also removed measurements where T was ≤ - 50⁰C or ≥ 100⁰C, or RH was > 99% as these corresponded to extreme conditions (~ 4.2% of all records). The remaining dataset contained 457,488,977 measurements from 11,933 sensors.

The 15 min data were averaged to 1 h intervals. A 75 % data completeness threshold was used (at least 3 15-minute measurements in an hour) based on channel A. This methodology ensured that the averages used were representative of hourly averages. We defined the hourly mean $PM_{2.5}$ cf_1 as the average of the $PM_{2.5}$ cf_1 measurements from channels A and B. We defined hourly mean $PM_{2.5}$ cf_atm as the average of $PM_{2.5}$ cf_atm measurements from channels A and B. We also calculated hourly mean T and RH from the 15-min averaged data from each PurpleAir sensor.

Overall, the dataset included 114,259,940 valid measurements with non-missing $PM_{2.5}$ data in Channels A or B corresponding to 11,932 PurpleAir sensors (8,312,155 measurements from 935 indoor sensors and 105,947,785 measurements from 10,997 outdoor sensors). A description of the number of sensors and measurements by state is provided in **Table S1** in *Supplementary Information*. (**Figure S1** in *Supplementary Information* displays the locations of indoor and outdoor PurpleAir sensors). Of the 11,932 PurpleAir sensors, 1,377 (~ 11.5%) had stopped reporting data at least a day before the data was downloaded (i.e., 20 July, 2021) while the remaining sensors were still in operation (**Figure 1**).

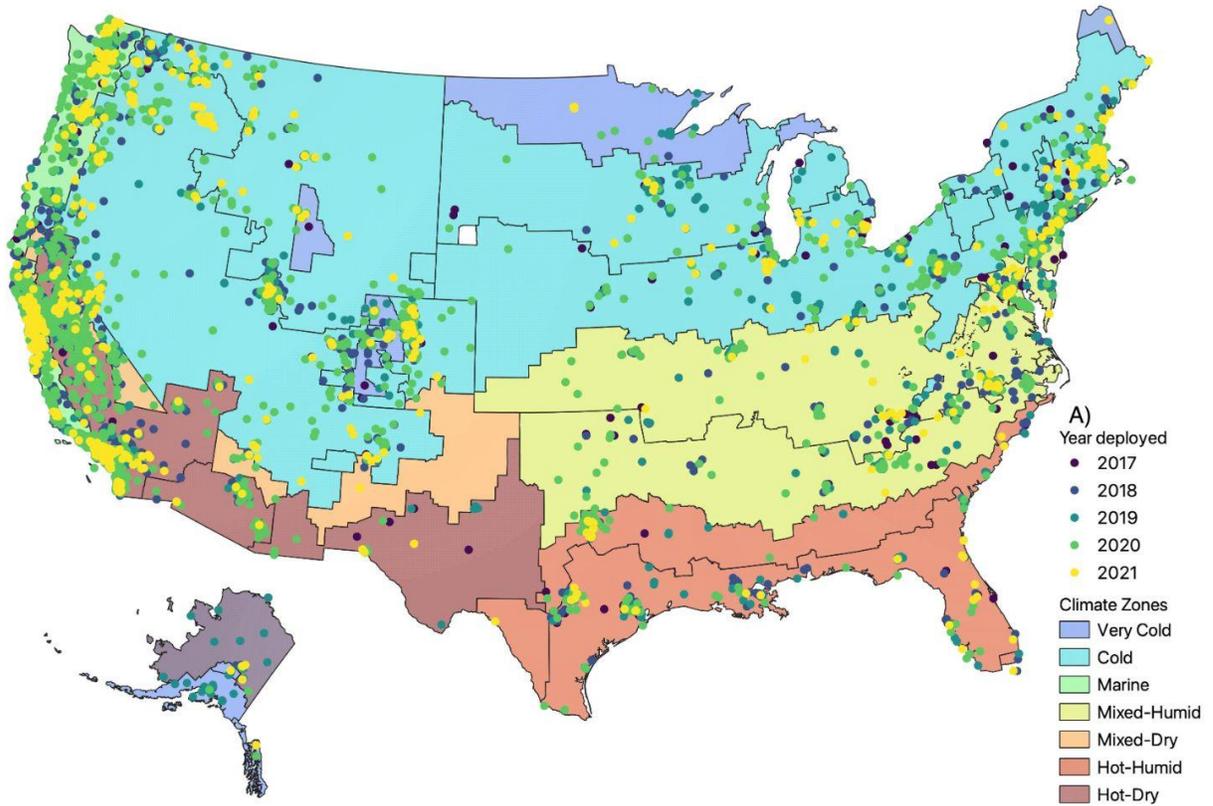

A)

Year deployed
- 2017
- 2018
- 2019
- 2020
- 2021

Climate Zones
- Very Cold
- Cold
- Marine
- Mixed-Humid
- Mixed-Dry
- Hot-Humid
- Hot-Dry

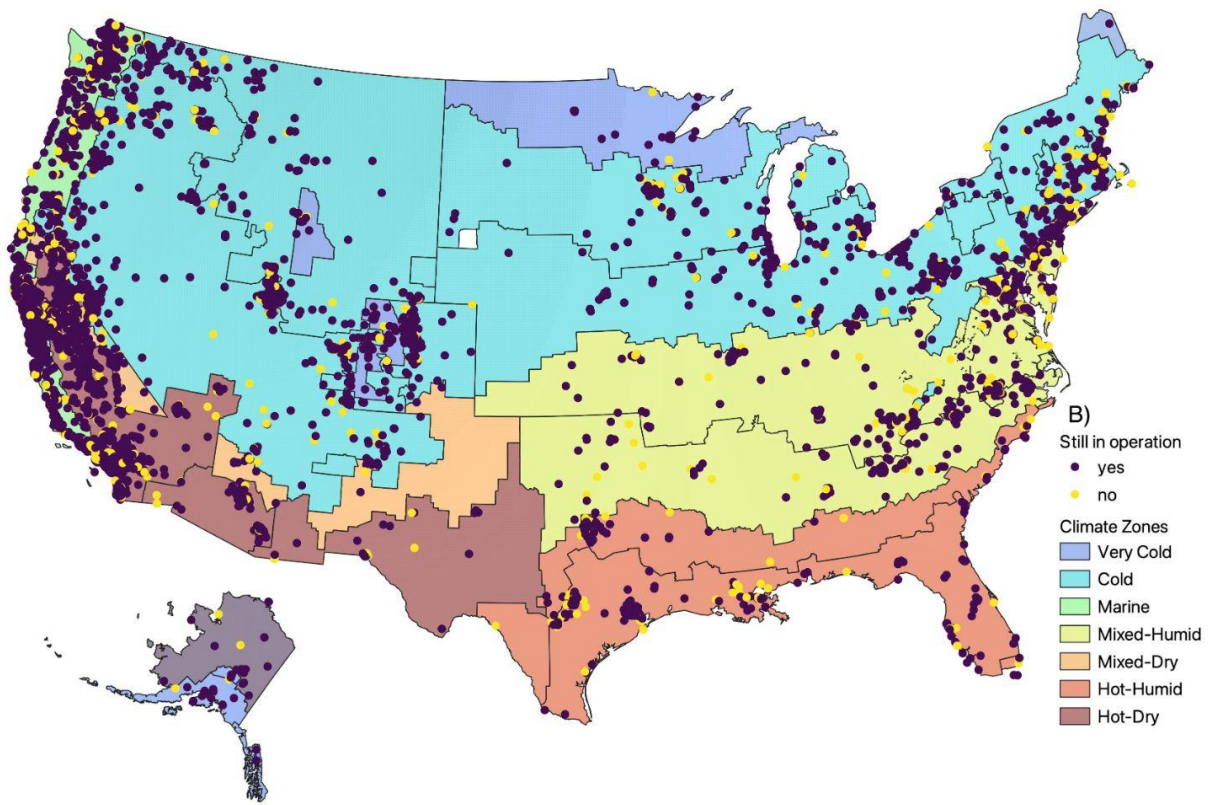

B)

Still in operation
- yes
- no

Climate Zones
- Very Cold
- Cold
- Marine
- Mixed-Humid
- Mixed-Dry
- Hot-Humid
- Hot-Dry

***Figure 1***: *The distribution of PurpleAir sensors considered in this analysis (Hawaii is not displayed) depicting (A) the year each sensor was deployed, and (B) If the sensor was removed before 20 July, 2021. Climate zones displayed are from the International Energy Conservation Code (IECC) Climate Zones ([https://codes.iccsafe.org/content/IECC2021P1/chapter-3-ce-general-requirements](https://codes.iccsafe.org/content/IECC2021P1/chapter-3-ce-general-requirements), last accessed August 31, 2022).*

## 2.2 Reference Measurements

Reference-grade (FRM/FEM) hourly PM$_{2.5}$ measurements between 1 January, 2017 and 20 July, 2021 were obtained from 80 EPA Air Quality System (AQS) regulatory monitoring sites (https://www.epa. gov/aqs, last accessed August 31, 2022) located within 50 meters from any outdoor PurpleAir sensor (**Table 1**). At eight of the sites (located in Indiana, Iowa, Michigan, Tennessee, Virginia and Washington) the monitoring method was updated midway during the period under consideration. Therefore, there were a total of 88 FRM/FEM monitors in our final analysis.

***Table 1***: *Location and type of the 88 reference PM$_{2.5}$ monitors within 50 meters of a PurpleAir sensor included in the current work. The number of merged PurpleAir and EPA measurements in each category is also listed.*

| Monitors | State |
|---|---|
| | |

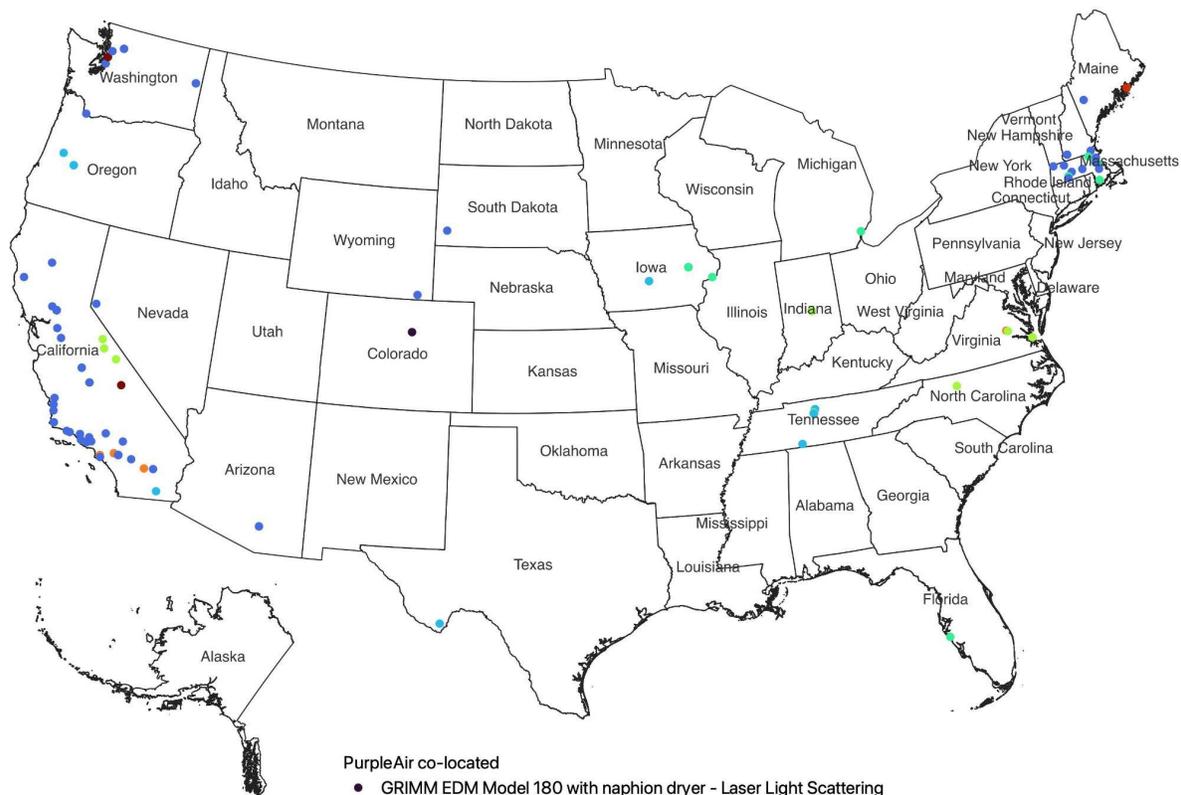

PurpleAir co-located

- ● GRIMM EDM Model 180 with naphion dryer - Laser Light Scattering
- ● Met One BAM-1020 Mass Monitor w/VSCC - Beta Attenuation
- ● Met One BAM-1022 Mass Monitor w/ VSCC or TE-PM2.5C - Beta Attenuation
- ● Teledyne T640 at 5.0 LPM - Broadband spectroscopy
- ● Teledyne T640X at 16.67 LPM - Broadband spectroscopy
- ● Thermo Scientific 1405-F FDMS w/VSCC - FDMS Gravimetric
- ● Thermo Scientific 5014i or FH62C14-DHS w/VSCC - Beta Attenuation
- ● Thermo Scientific Model 5030 SHARP w/VSCC - Beta Attenuation
- ● Thermo Scientific TEOM 1400 FDMS or 1405 8500C FDMS w/VSCC - FDMS Gravimetric

| | |
|---|---|
| 48 of the monitors in our sample were Met One BAM-1020 Mass Monitor w/VSCC - Beta Attenuation (1,002,533 merged measurements) | 33 reference monitors were in California (866,197 merged measurements) |
| 9 were Met One BAM-1022 Mass Monitor w/ VSCC or TE-PM2.5C - Beta Attenuation (218,084 merged measurements) | 16 in Massachusetts (26,930 merged measurements) |
| 10 were Teledyne T640 at 5.0 LPM - Broadband spectroscopy (97,706 merged measurements) | 9 in Washington (102,382 merged measurements) |
| 8 were Teledyne T640X at 16.67 LPM - Broadband spectroscopy (88,040 merged measurements) | 5 in Tennessee (88,505 merged measurements) |
| 6 were Thermo Scientific 5014i or FH62C14-DHS w/VSCC - Beta Attenuation (52,116 merged measurements) | 4 in Virginia (33,353 merged measurements) |
| 3 were Thermo Scientific TEOM 1400 FDMS or 1405 8500C FDMS w/VSCC - FDMS Gravimetric (21,591 merged measurements) | 4 in Iowa (199,138 merged measurements) |
| 2 were Thermo Scientific 1405-F FDMS w/VSCC - FDMS Gravimetric (15,872 merged measurements) | 2 in Maine (8,575 merged measurements) |
| 1 was GRIMM EDM Model 180 with naphion dryer - Laser Light Scattering (1,000 merged measurements) | 2 in Oregon (33,554 merged measurements) |
| 1 was a Thermo Scientific Model 5030 SHARP w/VSCC - Beta Attenuation monitor (3,199 merged measurements) | 2 in Indiana (26,499 merged measurements) |
| | 2 in Michigan (13,678 merged measurements) |
| | 1 each in Arizona (6,045 merged measurements) Colorado (1,000 merged measurements) Florida (15,434 merged measurements) Nevada (17,146 merged measurements) New Hampshire (30,591 merged measurements) North Carolina (27,253 merged measurements) South Dakota (1,879 merged measurements) Texas (364 merged measurements) Wyoming (1,618 merged measurements) |

## 2.3 Merging PurpleAir and Reference measurements

We paired hourly averaged $PM_{2.5}$ concentrations from 151 outdoor PurpleAir sensors that were within 50 meters from a reference monitor with the corresponding reference measurements. We removed records with missing EPA $PM_{2.5}$ data or where reference $PM_{2.5}$ measurements were < 0. The dataset contained a total of 1,500,141 merged concentrations with non-missing PurpleAir and EPA $PM_{2.5}$ values (**Table 1**).

If there was more than one reference monitor within 50 meters of a PurpleAir sensor, measurements were retained from one of the reference monitors. We prioritized retaining data from reference monitors that did not rely on light scattering techniques as these instruments tend to have additional error when estimating aerosol mass [28].

From the resulting dataset, we found that the Pearson correlation coefficient (R) between mean $PM_{2.5}$ cf_1 and reference $PM_{2.5}$ concentrations was 0.86, whereas the correlation between $PM_{2.5}$ cf_atm and reference $PM_{2.5}$ concentrations was 0.83. Henceforth, when describing PurpleAir measurements, we consider only the mean $PM_{2.5}$ cf_1 concentrations.

## 2.4 Evaluating Degradation

### 2.4.1 Method 1: 'Flagged' PurpleAir measurement

A flagged measurement, an indication of likely sensor degradation, is equal to a value of one when the channel A and B of the PurpleAir sensor differ. Barkjohn et al., (2021) defined a flagged measurement as one where the absolute difference between 24-hr averaged $PM_{2.5}$ from channel A and channel B ($\Delta$) > 5 µg/m$^3$ and the percent (%) difference between channels A and B: $\frac{abs(A-B) \times 2}{(A+B)}$ was > **2** standard deviations of the percentage difference between A and B for each PurpleAir sensor[26]. The absolute difference of 5 was chosen to avoid excluding too many measurements at low PM concentrations, whereas defining a threshold based on the % difference between channels A and B was chosen to avoid excluding too many measurements at high concentrations.

A data driven approach was adopted to determine if we should use a similar threshold in this study. We flagged measurements where the $\Delta$ > 5 µg/m$^3$ and when the % difference between channels A and B was greater than the top percentile of the distribution of the % difference between A and B channels for each PurpleAir sensor. We allowed the percentile threshold to range from 0.0 - 0.99, by increments of 0.01. We use percentiles as a threshold instead of standard deviation as the % difference between channels A and B is not normally distributed. At each step, we then compared the unflagged PurpleAir measurements with the corresponding reference data using the metrics: Pearson correlation coefficient (R) and the normalized root mean squared error (nRMSE). The percentile threshold that led to the best agreement between the PurpleAir sensor and the corresponding reference monitor was chosen. We calculated nRMSE in this study by normalizing the root mean square error (RMSE) by the standard deviation of $PM_{2.5}$ from the corresponding reference monitor. As a sensitivity test, we repeated the above analysis after removing records where the reference monitor relied on a light scattering technique (namely the Teledyne and the Grimm instruments), thus eliminating the more error-prone data (**Figure S3**). We note that past studies have shown that the Beta-Attenuation Mass Monitors (BAM) are likely to experience more noise at low $PM_{2.5}$ concentrations [28,29].

After determining the threshold to flag measurements using the collocated data (**Figure 2**), we evaluated the number of flagged measurements for each of the 11,932 PurpleAir sensors in our sample. We propose the percentage of flagged measurements at a given operational time (from the time, in hours, since each sensor started operating) as a potential degradation outcome. To visually examine if a threshold value existed beyond which these outcomes increased significantly, we plotted this outcome as well as the percentage of cumulative flagged measurements over time (**Figure 3)**. We evaluated whether the distribution of $PM_{2.5}$, RH and T conditions for flagged measurements is statistically different from that for unflagged measurements (**Table 2**).

For each PurpleAir sensor, at each operational hour, we evaluated the mean of all flags at the given hour and for all subsequent hours. We designated a PurpleAir sensor as permanently degraded if there were more than 100 hours where the cumulative mean of the 'flag' indicator of a given PurpleAir sensors was greater than or equal to 0.4 (i.e., 40% of subsequent measurements were degraded for at least 100 hours of operation) (**Figure 4; Figure S4**). In sensitivity analyses, we evaluated the number of PurpleAir sensors that would be considered 'degraded' for different thresholds (**Figure S5**). We also examined where such sensors were deployed.

A limitation of using the percentage of flagged measurements as a degradation metric is that it does not account for the possibility that channels A and B might both degrade in a similar manner. Therefore, we rely on a second approach, using collocated reference monitoring measurements, to evaluate this aspect of possible degradation.

## 2.4.2 Method 2: Evaluating the time-dependence of the error between corrected PurpleAir and reference measurements

PurpleAir data is often corrected using an algorithm to predict, as accurately as possible, the 'true' $PM_{2.5}$ concentrations based on reported PurpleAir concentrations. At the collocated sites, the reference $PM_{2.5}$ measurements, which are considered the true $PM_{2.5}$ concentrations, are the dependent variable in the models. Flagged PurpleAir measurements were first removed in the merged dataset (~2.5 % of all measurements: ~ 151 PurpleAir sensors). We then used the following Equation 1 proposed in Barkjohn et al., (2021)[26] to correct the PurpleAir sensors with the corresponding reference measurement:

$PM_{2.5, reference}$ = $PM_{2.5}$ x $s_1$ + RH x $s_2$ + b + ε  ……………………………………… **Equation (1)**

Where $PM_{2.5,reference}$ is the reference monitor measurement; $PM_{2.5}$ is the PurpleAir measurement calculated by averaging concentrations reported by channels A and B; RH is the relative humidity reported by the PurpleAir sensor. We empirically derived coefficients: $s_1$, $s_2$ and $b$ by regressing uncorrected PurpleAir $PM_{2.5}$ measurements on reference measurements of $PM_{2.5}$. ε denotes error from a standard normal distribution. We evaluated one correction model for all PurpleAir sensors in our dataset in a similar manner to Barkjohn et al., (2021). We evaluated and plotted the correction error which is defined as the difference between the corrected measurement and corresponding reference $PM_{2.5}$ measurement. In supplementary analyses, we repeat this process using 9 additional correction functions ranging from simple linear regressions to more complex machine learning algorithms (**Table S3**). A full description of these additional models can be found in deSouza et al., (2022) [18]. We chose to use this array of models, because some additionally correct the data for T as well as RH. In addition, research has shown that a non-linear correction equation might be more suitable to correct for PurpleAir measurements above ~ 500 µg/m$^3$ of PM$_{2.5}$ levels [30]. The machine learning models that we used in the supplement can identify such patterns using statistical learning.

## 2.5 Evaluating associations between the degradation outcomes and time

We evaluated the association between the degradation outcomes under consideration on time of operation using a simple linear regression (**Figure 5**):

$$Degradation\ Outcome\ =\ f\ +\ d\ \times\ hour\ of\ operation\ +\ \varepsilon\ ..............\ \textbf{Equation (2)}$$

where *f* denotes a constant intercept; *d* denotes the association between operational time (number of hours since each sensor was deployed) and the degradation outcome is the percentage of (cumulative) flagged measurements over all PurpleAir sensors at a given operational time; and $\varepsilon$ denotes error from a standard normal distribution.

For the degradation outcomes under consideration, we evaluated whether the associations were different in subgroups stratified by IECC Climate Zones that represent different T and RH conditions (**Table S2** contains information on PurpleAir measurements by climate zone). When evaluating the impact of climate zone on the percentage of flagged measurements, we examined the impact on outside devices alone, as indoor environments may not always reflect outside conditions due to heating, cooling, general sheltering, etc. Note that when joining climate zones with the complete dataset of PurpleAir IDs, there were a handful of sensors which did not fall within a climate zone. (This was not the case for our subset of collocated PurpleAir sensors.) We removed data corresponding to these sensors when evaluating climate zone-specific associations, corresponding to 2.9% of all data records (**Figure S2** in *Supplementary Information* shows where these sensors were located).

We also tested whether the cumulative number of PM$_{2.5}$ measurements recorded over 50, 100 and 500 µg/m$^3$ by individual PurpleAir sensors significantly modifies the association between operational time and the correction error, as previous work has found that low-cost optical PM sensors can degrade after exposure to high PM concentrations [21]. As the correction error will be larger at higher PM$_{2.5}$ concentrations [18,31], we also evaluated this association after normalizing the correction error by ($PM_{2.5, corrected} + PM_{2.5, reference}$)/2 to make it easier to interpret how cumulative exposure to high $PM_{2.5}$ measurements can affect the association between degradation and hour of operation.

The merged PurpleAir and reference measurements dataset only included measurements from outdoor PurpleAir sensors. We also evaluated the indoor/outdoor-specific associations between percentage flagged measurements and hour of operation.

Finally, we tested for potential non-linearities between the degradation outcomes under consideration and time of operation. Penalized splines (p-splines) were used to flexibly model the associations between the error and time of operation using a generalized additive model [GAM; degradation outcome ~ s(hour)]. We used a generalized cross-validation (GCV) criteria to select the optimal number of degrees of freedom (df) and plotted the relationships observed. Ninety-five percent confidence intervals (CIs) were evaluated by m-out-n bootstrap, which creates a non-parametric CIs by randomly resampling the data . Briefly, we selected a bootstrapped sample of monitors, performed the correction and then fit GAMs in each bootstrap sample using sensor ID clusters (100 replicates). (**Figure 6**).

All analyses were conducted using the software R. In all analyses, p-values < 0.05 were taken to represent statistical significance.

# 3 Results

## 3.1 Defining a 'flagged' PurpleAir measurement

**Figures 2a** and **2b** display agreement between the unflagged hourly PurpleAir measurements and the corresponding regulatory measurements using the R and nRMSE metrics, for different percentile thresholds to define a 'flag'. The lowest nRMSE and highest R were observed for the following definition of a flagged PurpleAir measurement: Absolute difference between $PM_{2.5}$ from channels A and B > 5 µg/m$^3$ and the % difference between channels A and B: $\frac{abs(A-B) \times 2}{(A+B)}$ > **85th percentile** of the percentage difference between channels A and B for each PurpleAir sensor. The 85th percentile of the percentage difference between channels A and B of each PurpleAir varies, with a mean of 38%. This definition resulted in about ~ 2% of the PurpleAir data being flagged (**Figure 2c**).

When we repeated this analysis excluding measurements from reference monitors that relied on light scattering techniques, using the 86th percentile yielded marginally better (the metrics differed by < 1%) results than using the 85th percentile (**Figure S3** in *Supplementary Information*). Given the small difference in results, the 85th percentile was used as the threshold in this study to define a flagged PurpleAir measurement.

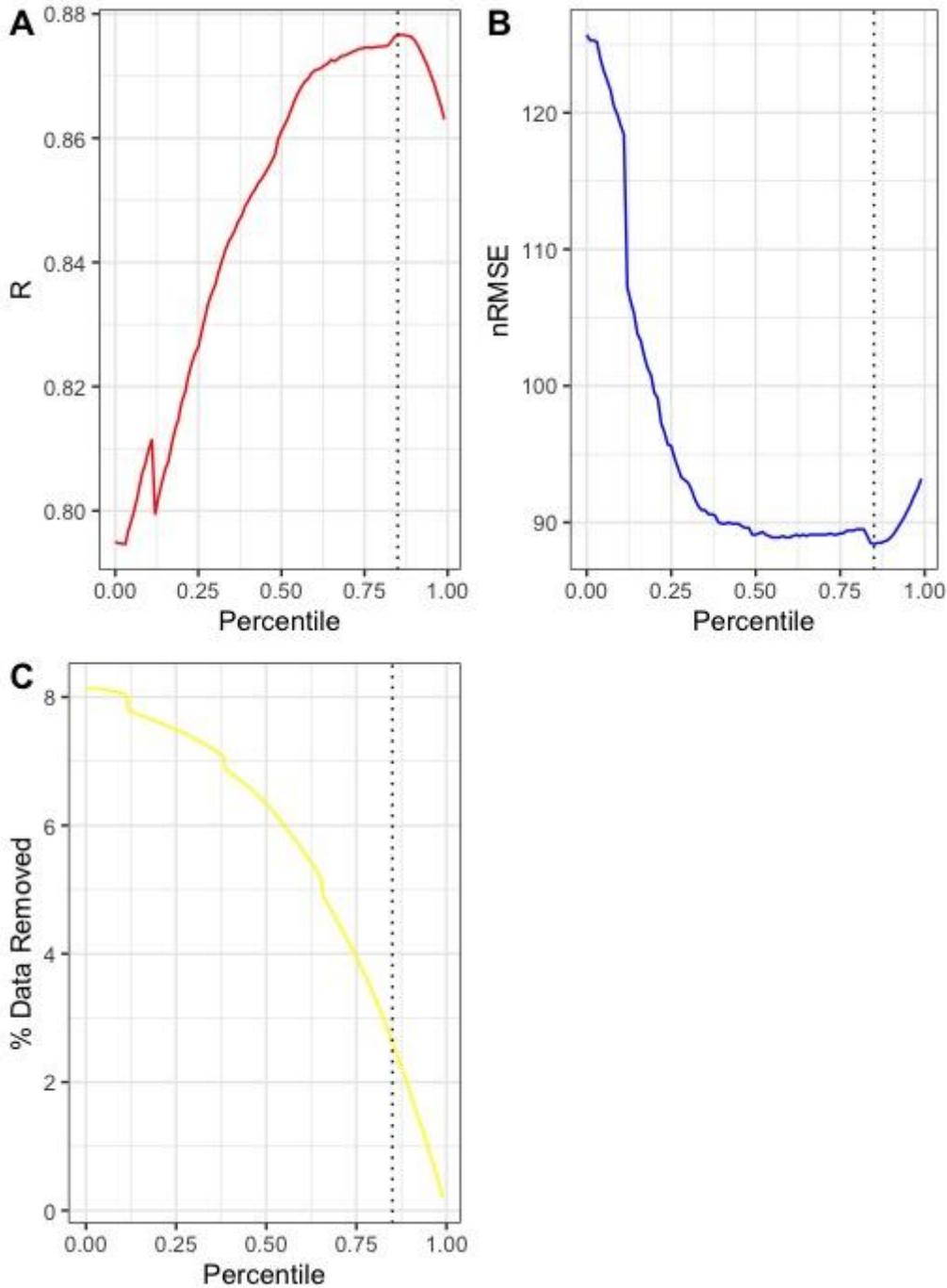

**Figure 2**: *Agreement between the hourly PurpleAir measurements and the corresponding reference measurements, where measurements are flagged and removed based on the criterion" | channel A - channel B | > 5 µg/m$^3$ and the % difference between channels A and B: $\frac{abs(A-B)\times2}{(A+B)} > x$ th percentile of the percentage difference between A and B for each PurpleAir sensor, where we vary $x$ between 0 - 0.99, captured by: A) Pearson correlation coefficient (R), and B) normalized root mean square error (nRMSE) metrics comparing unflagged measurements and the corresponding reference data based on different threshold percentile values. C) The % of measurements that were removed (because they were flagged) when evaluating R and nRMSE, for different percentile thresholds applied to the data are also displayed. The dotted vertical line represents the 85th percentile which corresponds to the lowest nRMSE and the highest R.*

## 3.2 Visualizing the degradation outcomes: Percentage of flagged measurements over time

Using the empirically derived definition of flagged measurements, the percentage of flagged measurements, as well as the percentage of cumulative flagged measurements across the 11,932 PurpleAir sensors for every hour of operation are plotted in **Figure 3**. The total number of measurements made at every hour of operation is also displayed using the right axis. The percentage of flagged measurements increases over time. After 4 years (~ 35,000 hours) of operation the percentage of flagged measurements every hour is ~ 4%. After 4 years of operation, we observe a dramatic increase in the average percentage of flagged measurements likely due to the small number of PurpleAir sensors operational for such long periods of time in our dataset. We also observe a high percentage of flagged measurements during the first 20 hours of the operation of all sensors.

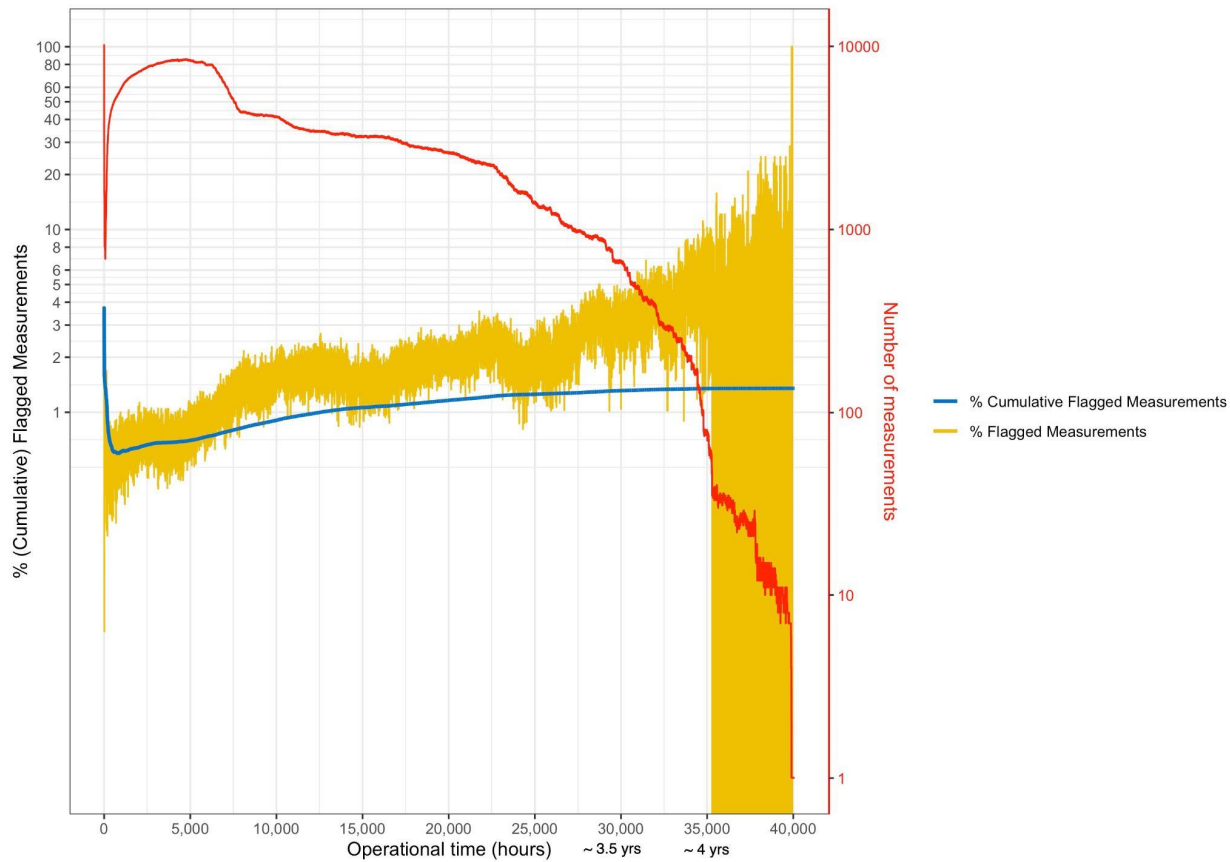

***Figure 3***: *Percentage of flagged PurpleAir measurements (yellow) and percentage cumulative flagged (blue) measurements at a given operational time (time since each sensor started operation in hours) as well as the number of measurements recorded (red) plotted on the secondary y-axis on the right over all the PurpleAir sensors considered in this analysis.*

Using *t*-tests we find that the mean of $PM_{2.5}$, T, and RH measurements were statistically different ($p < 0.05$) for flagged PurpleAir measurements compared to unflagged measurements (**Table 2**). $PM_{2.5}$ and T measurements recorded when a measurement was flagged were higher than for unflagged measurements, whereas RH tended to be lower. The differences between RH and T values for flagged versus non-flagged measurements are small. The difference in $PM_{2.5}$ distribution was due in part to the way flags have been defined. As data are flagged only if concentrations are at least 5 µg/m$^3$ different, the minimum average flagged concentration is 2.5 µg/m$^3$ (e.g., A=0, B=5). There are no notable differences between the percentage of flagged measurements made every month.

**Table 2**: PM$_{2.5}$, temperature and RH values, and months corresponding to flagged and unflagged measurements.

| | Unflagged Data (n=112,716,535, 99%) | Flagged Data (n= 1,543,405, 1%) |
|---|---|---|
| **Raw Mean PM$_{2.5}$** (Mean of Channel A and Channel B) (µg/m$^3$) | Min/Max: 0/1459<br>Mean: 10<br>Median: 5<br>1st quartile: 2<br>3rd quartile: 11 | Min/Max: 2.5/1339<br>Mean: 26<br>Median: 14<br>1st quartile: 7<br>3rd quartile: 27 |
| **RH (%)** | Min/Max: 0/99<br>Mean: 46<br>Median: 48<br>1st quartile: 34<br>3rd quartile: 59 | Min/Max: 0/99<br>Mean: 43<br>Median: 44<br>1st quartile: 30<br>3rd quartile: 57 |
| **Temperature ($^0$C)** | Min/Max: -42/68<br>Mean: 18<br>Median: 18<br>1st quartile: 11<br>3rd quartile: 24 | Min/Max: -46/89<br>Mean: 19<br>Median: 19<br>1st quartile: 13<br>3rd quartile: 26 |
| **Month** | Jan: 10,233,928 (98.5%)<br>Feb: 9,650,954 (98.4%)<br>March: 10,979,861 (98.7%)<br>April: 10,989,824 (98.9%)<br>May: 11,671,186 (98.8%)<br>June: 11,674,808 (98.6%)<br>July: 9,555,217 (98.6%)<br>Aug: 5,246,854 (98.7%)<br>Sep: 6,248,360 (98.6%)<br>Oct: 8,025,096 (98.8%)<br>Nov: 8,759,251 (98.6%)<br>Dec: 9,681,196 (98.5%) | Jan: 157,728 (1.5%)<br>Feb:156,615 (1.6%)<br>March: 141,003 (1.3%)<br>April: 125,060 (1.1%)<br>May: 143,421 (1.2%)<br>June: 160,317 (1.4%)<br>July: 140,255 (1.4%)<br>Aug: 67,196 (1.3%)<br>Sep: 86,200 (1.4%)<br>Oct: 99,753 ((1.2%)<br>Nov: 120,721 (1.4%)<br>Dec: 145,136 (1.5%) |

We next evaluated the number of PurpleAir measurements that were permanently degraded, or that had a cumulative mean of flag over subsequent hours of operation ≥ 0.4 for at least 100 hours of operation (i.e., at least 40% of measurements flagged) (**Figure 4**). **Table 3** displays the fraction of permanently degraded sensors in different climate zones and different locations (inside/outside). It appears that the largest fraction of degraded sensors occurred in the south-east United States, a hot and humid climate. **Figure S4** displays the cumulative mean of flag for each 'permanently degraded' sensor (the title of each plot corresponds to the sensor ID as provided on the PurpleAir website) at each instance of time. **Figure S4** also depicts the starting year of each permanently degraded sensor. The sensor age varied widely over the set of permanently degraded sensors, indicating that permanent degradation is not time-related.

Note that from **Figure S4** some of the 240 sensors identified appear to recover or behave normally after a long interval of degradation (cumulative mean of flag decreases). This could be an artifact of the way the cumulative mean of the flagged indicator is calculated. If the final few measurements of the sensor are not flagged, then the cumulative mean for the final hours of operation of the PurpleAir sensors might be low. It is also possible that some of the sensors could have been temporarily impacted by dust or insects. The owner of the PurpleAir sensors might have replaced the internal Plantower sensors or cleaned out the sensors which could have caused the sensors to recover.

**Figure S5A** and **S5B** are maps showing locations of PurpleAir sensors that had cumulative mean of 'flag' over subsequent hours of operation of ≥ 0.3 (number of sensors = 323) and 0.5 (number of sensors = 182), respectively for at least 100 hours of operation.

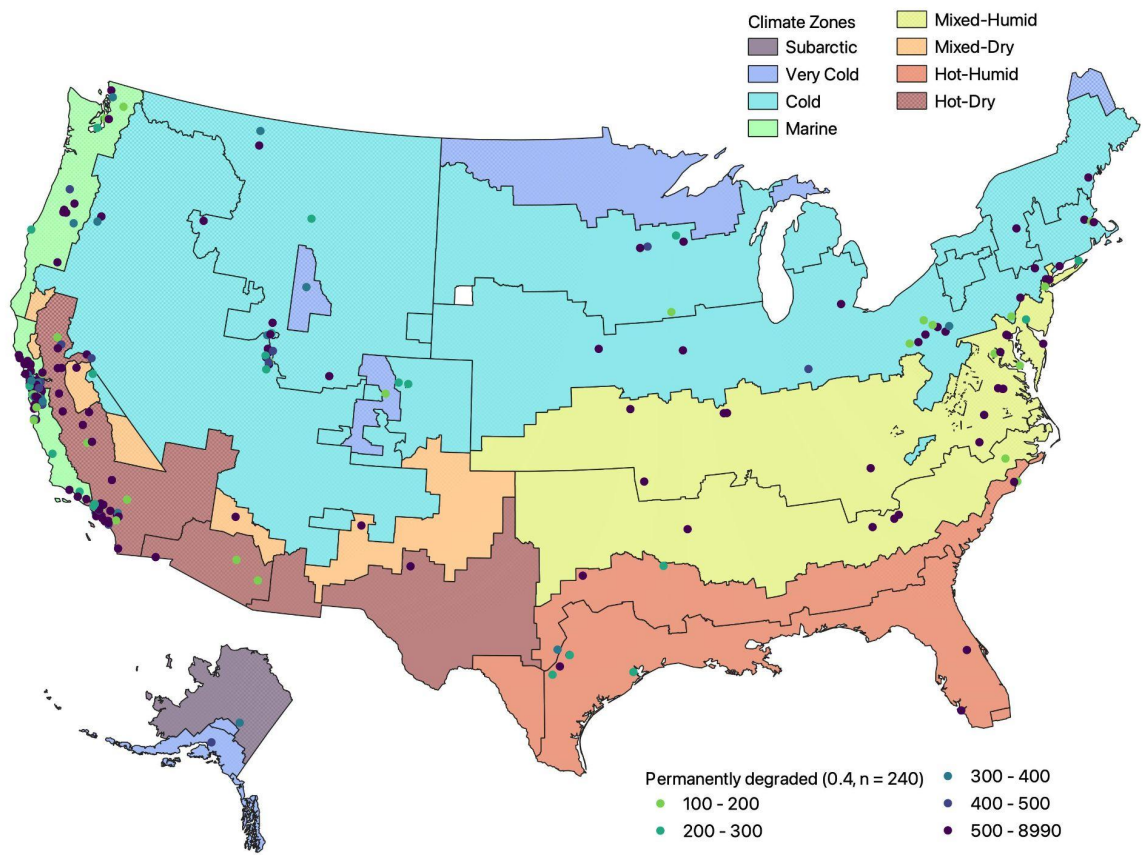

***Figure 4:*** *Map of permanently degraded PurpleAir sensors with at least 100 measurements for which the cumulative mean of the flagged indicator ≥ 0.4. The number of hours of operation for which the cumulative mean of the flag indicator is ≥ 0.4 is indicated by point color*

**Table 3**: *Fraction of permanently degraded PurpleAir sensors in climate zones and locations*

|  | Percentage of permanently degraded sensors |
|---|---|
| **All** | 240 out of 11,932 (2.0%) |
| **Device Location** | |
| **Inside** | 2 out of 935 (0.21%) |
| **Outside** | 238 out of 10,997 (2.2%) |
| **Climate Zone** | |
| **Cold** | 51 out of 2,458 (2.1%) |
| **Hot-Dry** | 54 out of 2,680 (2.0%) |
| **Hot-Humid** | 11 out of 281 (3.9%) |
| **Marine** | 84 out of 4,842 (1.7%) |
| **Mixed-Dry** | 3 out of 361 (0.8%) |
| **Mixed-Humid** | 24 out of 750 (3.2%) |
| **Sub Arctic** | 1 out of 58 (1.7%) |
| **Very Cold** | 3 of 108 (2.8%) |
| **No information** | 9 of 394 (2.3%) |

## 3.3 Visualization of the error in the corrected PurpleAir PM$_{2.5}$ measurements over time

The correction function derived using a regression analysis yielded the following function to derive corrected PM$_{2.5}$ concentrations from the raw PurpleAir data: PM$_{2.5,corrected}$ = 5.92 + 0.57PM$_{2.5,raw}$ -0.091RH. After correction, the Pearson correlation coefficient (R) improved slightly, from 0.88 to 0.89, and the RMSE improved significantly, from 12.5 to 6.6 µg/m$^3$. The mean, median and maximum error observed were 3.3, 2.2, and 792.3 µg/m$^3$, respectively (**Table S3**). **Figure 5** displays the mean correction error across all sensors for every hour in operation. The error becomes larger after 35,000 hours (3 years). We note that similar results were observed when using a wide array of correction models, including models that contain both RH and T as variables, as well as more complex machine learning models that yielded the best correction results (Random Forest: R=0.99, RMSE = 2.4 μg/m$^3$) (**Table S3**).

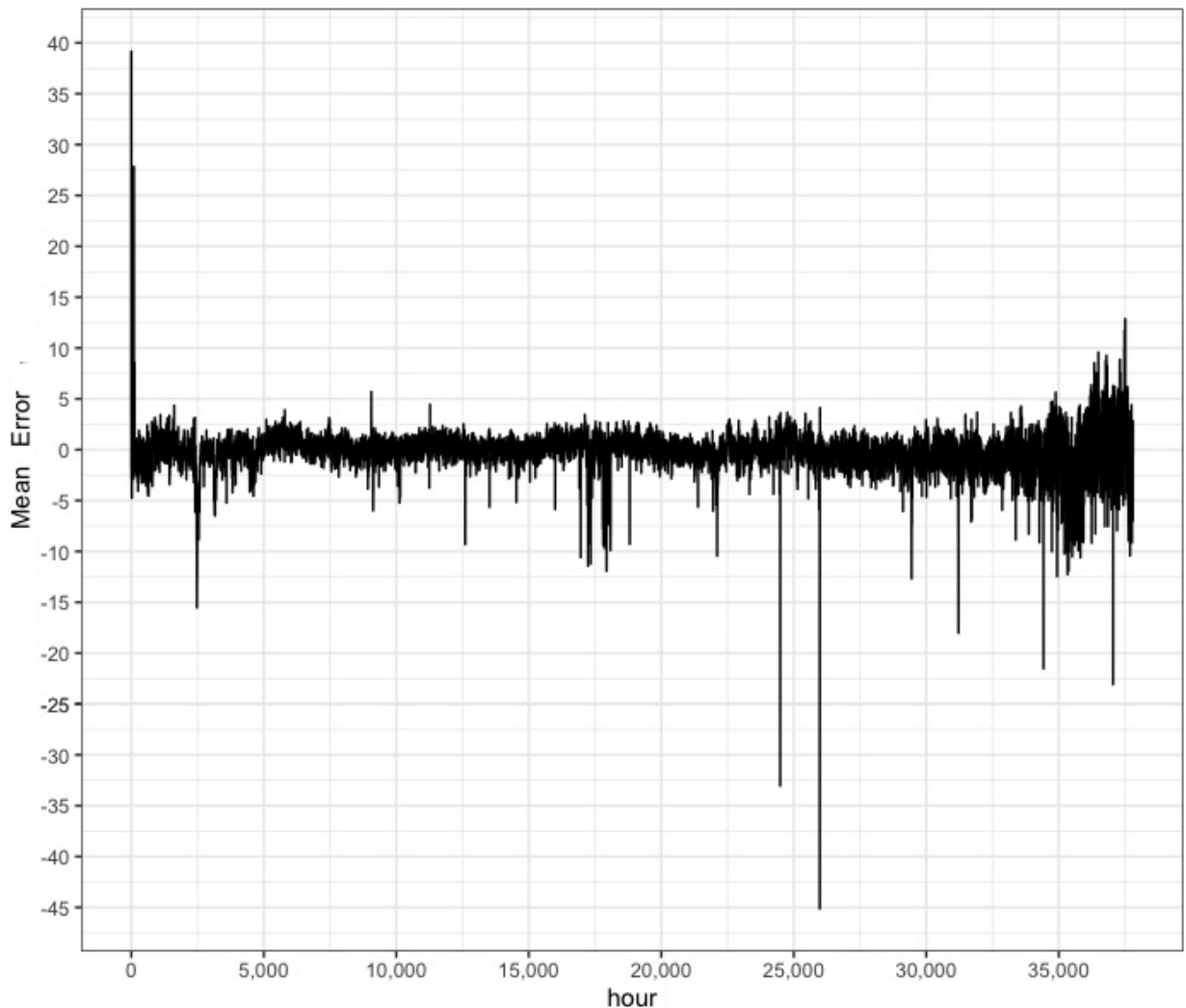

**Figure 5**: *Mean difference between the corrected PM$_{2.5}$ measurements from the PurpleAir sensors and the corresponding reference PM$_{2.5}$ measurements across all sensors as a function of hour of operation.*

## 3.4 Associations between degradation outcomes and operational times

We assessed the association between degradation outcomes and operational time based on Equation 2. We observed that the percentage of flagged measurements increased on average by 0.93% (95% CI: 0.91%, 0.94%) for every year of operation of a PurpleAir sensor. Device location and climate zone were significant effect modifiers of the impact of time-of-operation on this degradation outcome. PurpleAir sensors located outside had an increased percentage of flagged measurements every year corresponding to 1.06% (95% CI: 1.05%, 1.08%), whereas those located inside saw the percentage of flagged measurements decrease over time. Outdoor PurpleAir sensors in hot-dry climates appeared to degrade the fastest with the percentage of flagged measurements increasing by 2.09% (95% CI: 2.07%, 2.12%) every year in this climate zone (**Table 3**). Hot-dry places are dustier. Dust can degrade fan performance which would lead to potentially more disagreement between channels A and B of the PurpleAir sensors.

The correction error ($PM_{2.5, corrected}$ - $PM_{2.5, reference}$) appeared to become negatively biased over time: -0.12 (95% CI: -0.13, -0.11) µg/m$^3$ per year of operation, except for sensors in hot and dry environments where the error was positively biased and increases over time by 0.08 (95% CI: 0.06, 0.09) µg/m$^3$ per year of operation. PurpleAir sensors cannot 'see' dust or larger particles, which dominate in hot-dry environments, potentially explaining the disagreement between the corrected PurpleAir and reference measurements in these environments. The magnitude of the correction error bias over time appears to be highest in hot and humid environments corresponding to -0.92 (95% CI: -0.11, -0.75) µg/m$^3$ per year. RH has a major impact on PurpleAir performance, so it is not altogether surprising that degradation appears to be highest in hot and humid environments. We observed similar results when regressing the correction errors derived using other correction model forms (**Table S4**). Climate zone is a significant modifier of the association between both degradation outcomes and time (**Table 4**).

The cumulative number of $PM_{2.5}$ measurements recorded over 50, 100 and 500 µg/m$^3$ significantly negatively modifies the association between operational time and the correction error (**Table S5**) meaning that sensors that experience more high concentration episodes are more likely to underestimate $PM_{2.5}$. The increase in the negative bias of the corrected sensor data could be because the absolute magnitude of the correction error will be higher in high $PM_{2.5}$ environments. When we evaluated the impact of the cumulative number of high $PM_{2.5}$ measurements on the association between the normalized correction error and operation hour (hours since deployment), we found that the cumulative number of high $PM_{2.5}$ measurements was not a significant effect modifier of this association (**Table S6**). In other words, we did not observe sensors in higher $PM_{2.5}$ environments degrading faster.

***Table 4***: *Associations between the degradation outcomes (% of flagged measurements and correction error) and year of operation of the PurpleAir sensors. Note that we did not have any PurpleAir sensors collocated with a regulatory monitor in Sub Arctic and Cold Climates. In addition, all PurpleAir monitors collocated with regulatory monitors were outdoor.*

| | Associations (95% Confidence Interval) |
| --- | --- |

| Dataset | Percentage of Flagged Measurements | Correction Error |
|---|---|---|
| **All** | 0.93* (0.91, 0.94) | -0.12* (-0.13, -0.11) |
| **Device Location** | | |
| **Inside** | -0.10* (-0.12, -0.09) | - |
| **Outside** | 1.06* (1.05, 1.08) | - |
| **Climate Zone (Outside Devices Only)** | | |
| **Cold** | 0.74* (0.71, 0.76) | -0.27* (-0.29, -0.25) |
| **Hot-Dry** | 2.09* (2.07, 2.12) | 0.08* (0.06, 0.09) |
| **Hot-Humid** | 0.35* (0.32, 0.37) | -0.92* (-0.11, -0.75) |
| **Marine** | 0.41* (0.39, 0.44) | -0.13* (-0.15, -0.10) |
| **Mixed-Dry** | -0.05* (-0.08, -0.02) | -0.31* (-0.40, -0.21) |
| **Mixed-Humid** | 0.54* (0.51, 0.57) | -0.28* (-0.33, -0.23) |
| **Sub Arctic** | -0.18* (-0.22, -0.14) | - |
| **Very Cold** | 0.13* (0.10, 0.16) | - |

(*p<0.05)

## 3.5 Evaluating potential non-linearities between the degradation outcomes and time

GCV criteria revealed that the dependence of the percentage of flagged PurpleAir measurements over time was non-linear, likely due to the non-linear relationship observed at operational times greater than 30,000 hours (3.5 years; **Figure 6**). However, due to the small number of measurements after this time interval, the shape of the curve after this time was uncertain, as evidenced by the wide confidence bands in this time period. The correction error appeared to become more and more negatively biased after 30,000 operational hours (3.5) years. However, due to the small number of sensors operating for more than 3 years, the wide confidence interval bands past 3 years casts uncertainty on the latter finding.

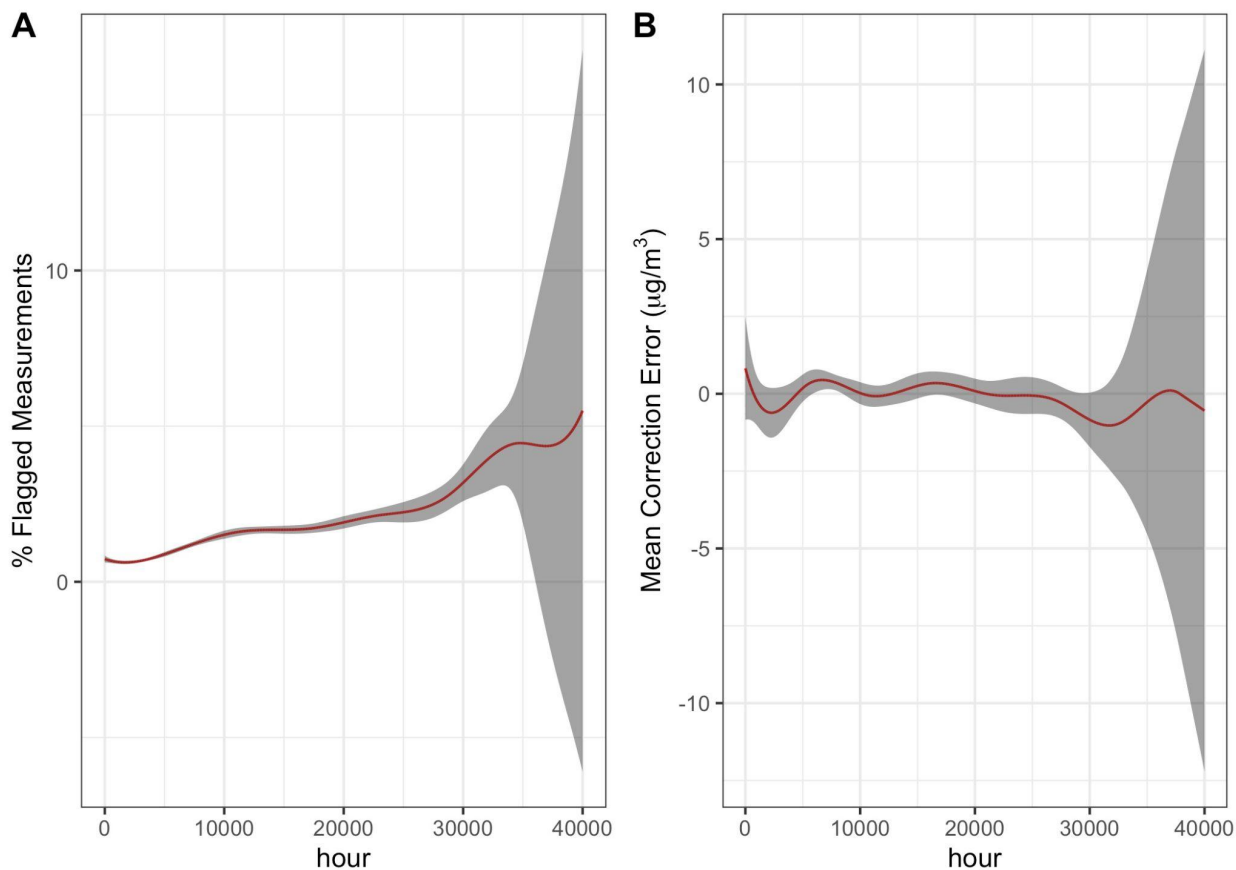

***Figure 6****: Response plot and 95% confidence intervals (shaded region) for the association between the degradation outcomes of (A) Percentage (%) of Flagged Measurements and (B) Correction Error with respect to operational time in hours generated using GAMs*

# 4 Discussion and Conclusions

We evaluated two proposed degradation outcomes for the PurpleAir sensors over time. We observed there were a large number of measurements from channels A and B of each sensor during the first 20 hours of operation that were flagged (**Figure 1**). Some of these data might come from lab testing of the PurpleAir sensors. Our results suggest that it is important to delete the first 20 hours of data when analyzing PurpleAir measurements. We observed that the percentage of flagged measurements (where channels A and B diverged) increased linearly over time and was on average ~4% after 3 years of operation. It appeared that measurements from PurpleAir sensors are fairly robust, at least during this period. There were only a small number of PurpleAir sensors (< 100) operational for over 3 years in the current dataset. Further work is needed to evaluate the performance of these devices after 3 years of operation.

Flagged measurements were more likely to be observed at higher $PM_{2.5}$ concentrations, lower RH levels and higher T levels (**Table 1**). When we evaluated associations between the percent of flagged measurements and year of operation for sensors in different locations (i.e. outdoor and indoor), we found that outdoor sensors degrade much faster than indoor sensors (**Table 3**). As T and RH impact the likelihood of observing a flagged measurement, this could be because environmental conditions of indoor environments (T and RH) are more regulated that outdoor environments, and indoor instruments tend to be more protected. Our results indicate that the percent of flagged measurements for indoor environments decrease over time. This could be because of the high percent of flagged measurements observed in the first 20 h of operation, and the lack of large changes in the percent of flagged measurements in later hours of operation in comparison to outdoor sensors. We also note that there is a much smaller number of indoor sensors in comparison to outdoor instruments (935 compared to 10,997), and thus far fewer measurements available at long operational time intervals.

For outdoor sensors, we found that the climate zone in which the sensor was deployed is an important modifier of the association between the % of flagged measurements and time. Outdoor sensors in hot-dry climates degrade the fastest, with the percentage of flagged measurements increasing by 2.09% (95% CI: 2.07%, 2.12%) every year, an order of magnitude faster than any other climate zone (**Table 3**).

There was a small number of PurpleAir sensors (240 out of 11,932) that were permanently degraded (the cumulative mean of subsequent measurements had over 40% degraded measurements for at least 100 hours). The list of permanently degraded PurpleAir IDs is presented in **Figure S4**. These sensors should be excluded when conducting analyses. The largest fraction of permanently degraded PurpleAir sensors appeared to be in the hot and humid climate zone indicating that sensors in these climates likely needed to be replaced sooner than in others (**Table 2**). There was no significant relationship between sensor age and permanent degradation, indicating that there may be other factors responsible for causing permanent failure among the PurpleAir sensors. For example, anecdotal evidence suggests that the PurpleAir sensors can be impacted by dust or even insects and degrade the internal components of one or the other PurpleAir channels.

When evaluating the time-dependence of the correction error, we found that the PurpleAir instrument bias changes by -0.12 (95% CI: -0.13, -0.11) $\mu g/m^3$ per year of operation. However, the low associations indicate that this bias is not of much consequence to the operation of PurpleAir sensors. Climate zone was a significant effect modifier of the association between bias and time. The highest associations were observed in hot and humid regions corresponding to -0.92 (95% CI: -0.11, -0.75) μg/m$^3$ per year. Exposure to a cumulative number of high $PM_{2.5}$ measurements significantly affected the association between the normalized correction error over time.

It is not altogether surprising that the correction error increases most rapidly in hot and humid climate zones, as past evidence suggests that the performance of PurpleAir are greatly impacted by RH. It is surprising that this is not the case for the other degradation outcomes considered in this study: % of flagged measurements. It is likely that this outcome increases most rapidly over time in hot and dry environments instead, because such environments also tend to be dusty. Dust can enter PurpleAir devices, leading to disagreement between the two Plantower sensors.

When accounting for non-linearities in the relationship between the correction error and time, **Figure 6a** indicates that the bias in the correction error is not linear; rather it increases significantly after 30,000 hours or 3.5 years. Overall, we found that more work is needed to evaluate degradation in PurpleAir sensors after 3.5 years of operation. Importantly, the degradation outcomes derived in this paper can be used to remove 'degraded' PurpleAir measurements in other analyses. We also show that concerns about degradation are more important in some climate zones than others, which may necessitate appropriate maintenance/cleaning procedures for sensors in different locations.

# Acknowledgements

The authors are grateful to Mike Bergin for several useful discussions. Thank you to PurpleAir (https://www2.purpleair.com/) for providing publicly the data that made this paper possible.

# Disclaimer

The views expressed in this paper are those of the author(s) and do not necessarily represent the views or policies of the US Environmental Protection Agency. Any mention of trade names, products, or services does not imply an endorsement by the US Government or the US Environmental Protection Agency. The EPA does not endorse any commercial products, services, or enterprises.

# Supplementary Information:

An analysis of degradation in low-cost particulate matter sensors

Priyanka deSouza[1,2*], Karoline Barkjohn[3], Andrea Clements[3], Jenny Lee[4], Ralph Kahn[5], Ben Crawford[6], Patrick Kinney[7]

1: Department of Urban and Regional Planning, University of Colorado Denver, Denver CO, 80202, USA
2: CU Population Center, University of Colorado Boulder, Boulder CO, 80302, USA
3: Office of Research and Development, US Environmental Protection Agency, 109 T.W. Alexander Drive, Research Triangle Park, NC 27711, USA
4: Department of Biostatistics, Harvard T.H. Chan School of Public Health, Boston, MA 02115, USA
5: NASA Goddard Space Flight Center, Greenbelt, MA, 20771, USA
6: Department of Geography and Environmental Sciences, University of Colorado Denver, 80202, USA
7: Boston University School of Public Health, Boston, MA, 02118 USA

*: Corresponding author priyanka.desouza@ucdenver.edu

# Figures

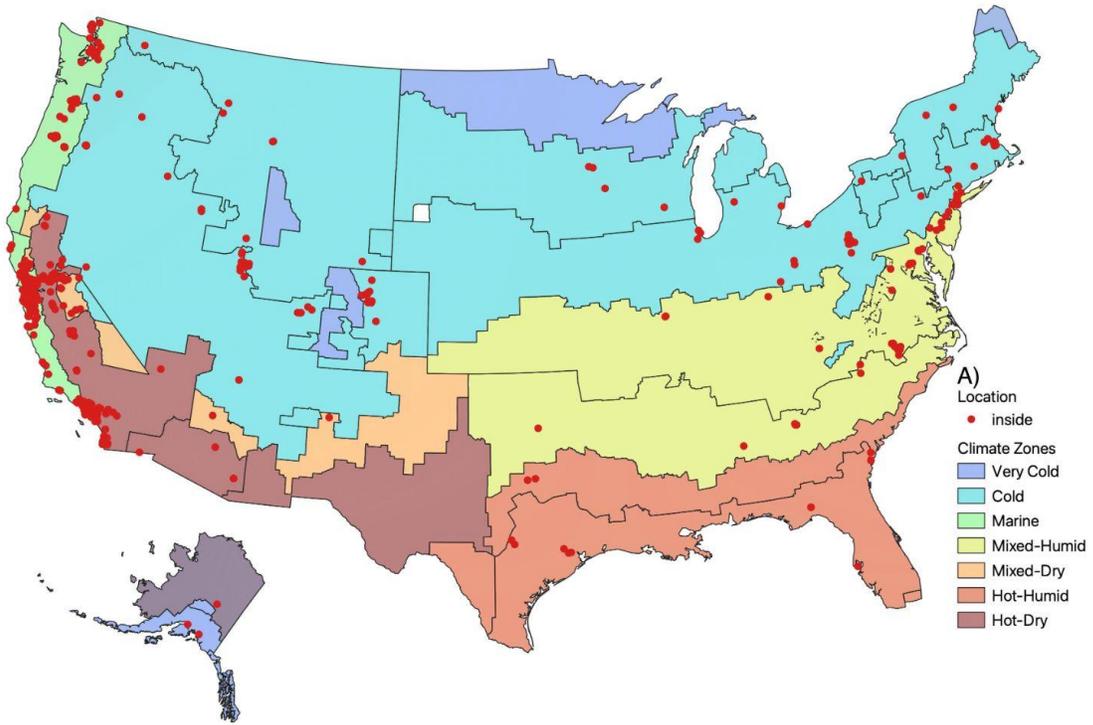

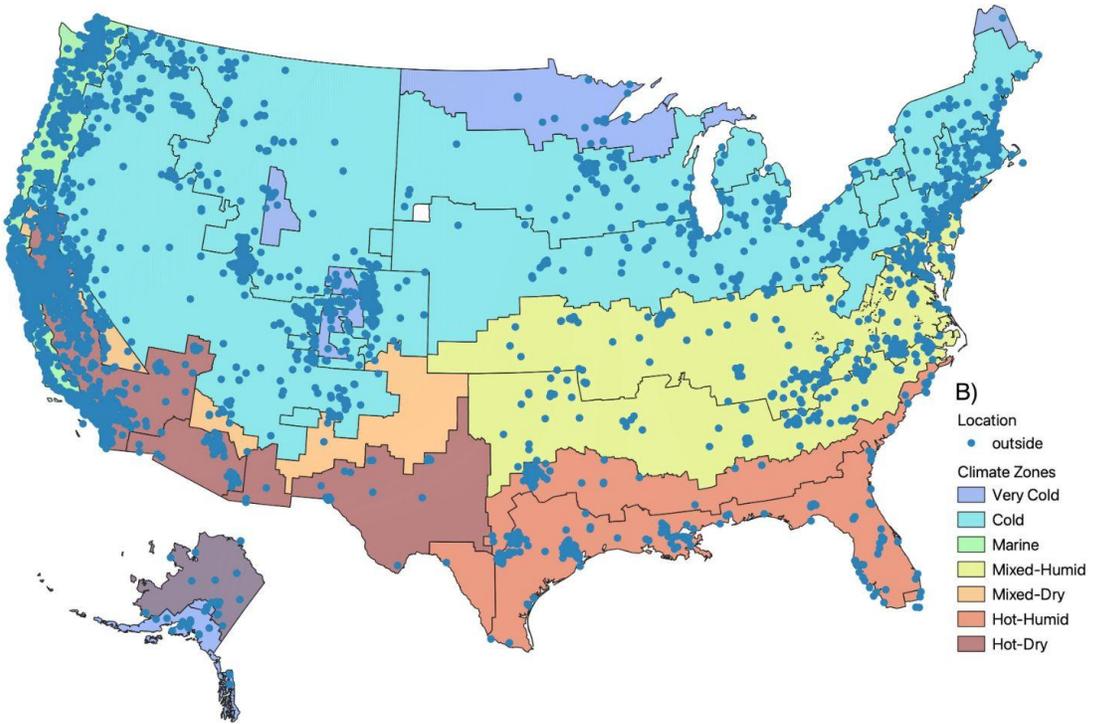

**Figure S1:** *A) Indoor and B) Outdoor PurpleAir monitors in the United States (Hawaii is not displayed) considered in this study*

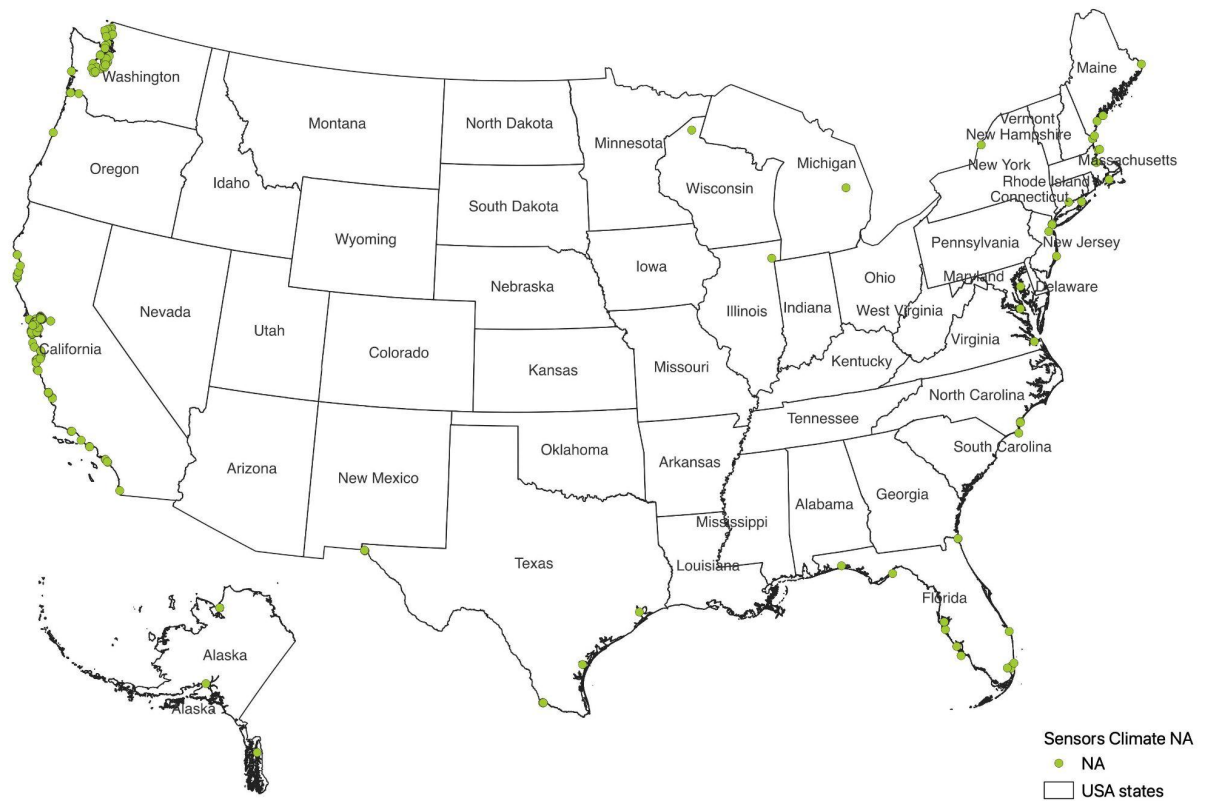

**Figure S2**: *Sensors for which we did not have climate information*

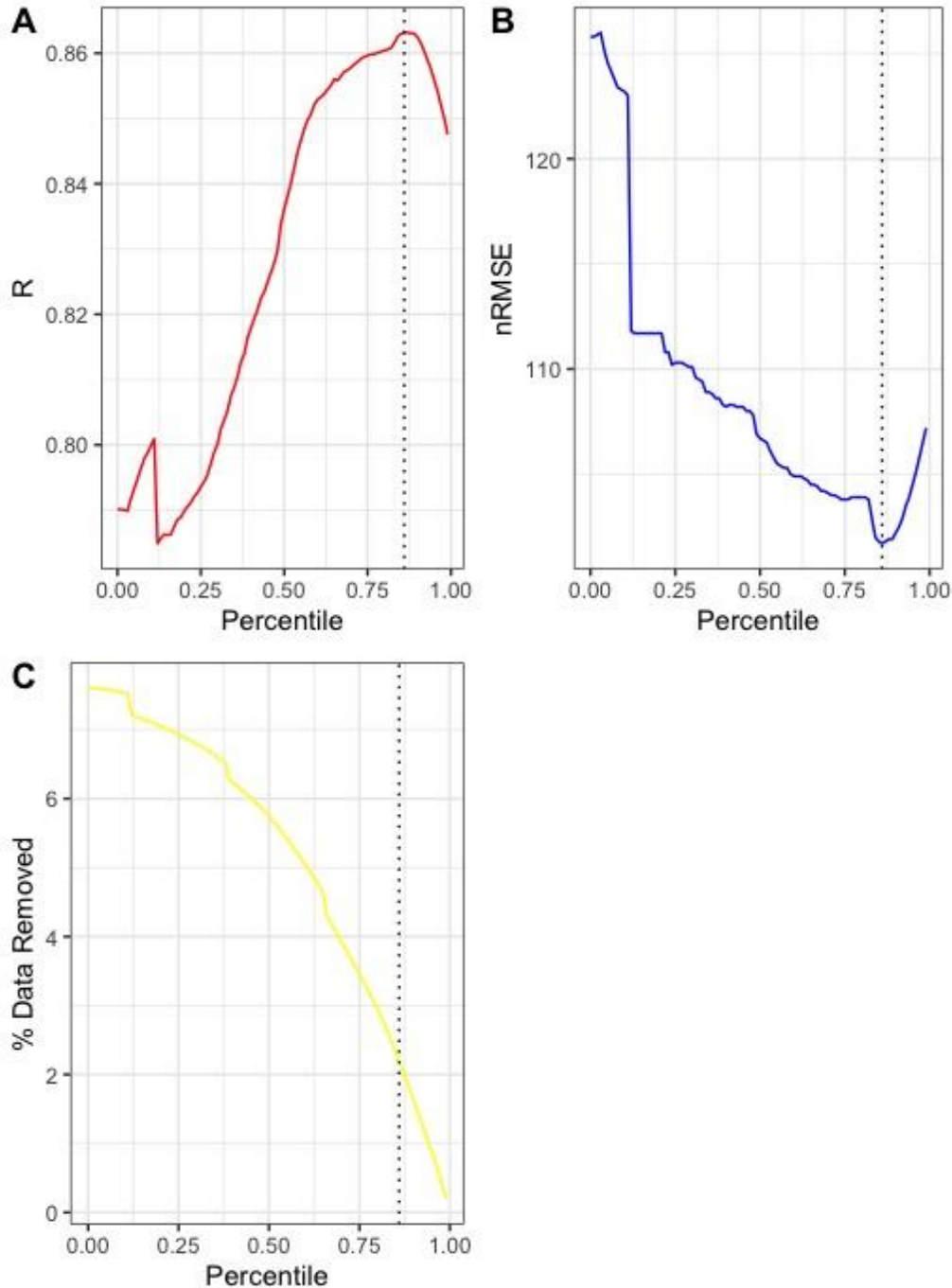

**Figure S3**: *Agreement between the hourly PurpleAir measurements and the corresponding reference measurements from instruments not relying on the principles of light scattering, where measurements are flagged and removed based on the criterion"| channel A - channel B | < 5 µg/m³ and the % difference between channels A and B: $\frac{abs(A-B)\times2}{(A+B)} > x$ th percentile of the percentage difference between A and B for each PurpleAir sensor, where we vary $x$ between 0 - 0.99, captured by: A) Pearson correlation coefficient (R), and B) normalized root mean square error (nRMSE) metrics*

*comparing unflagged measurements and the corresponding reference data based on different threshold percentile values. C) The % of measurements that were removed (because they were flagged) when evaluating R and nRMSE, for different percentile thresholds applied to the data are also displayed. The dotted vertical line represents the 86th percentile which corresponds to the lowest nRMSE and the highest R.*

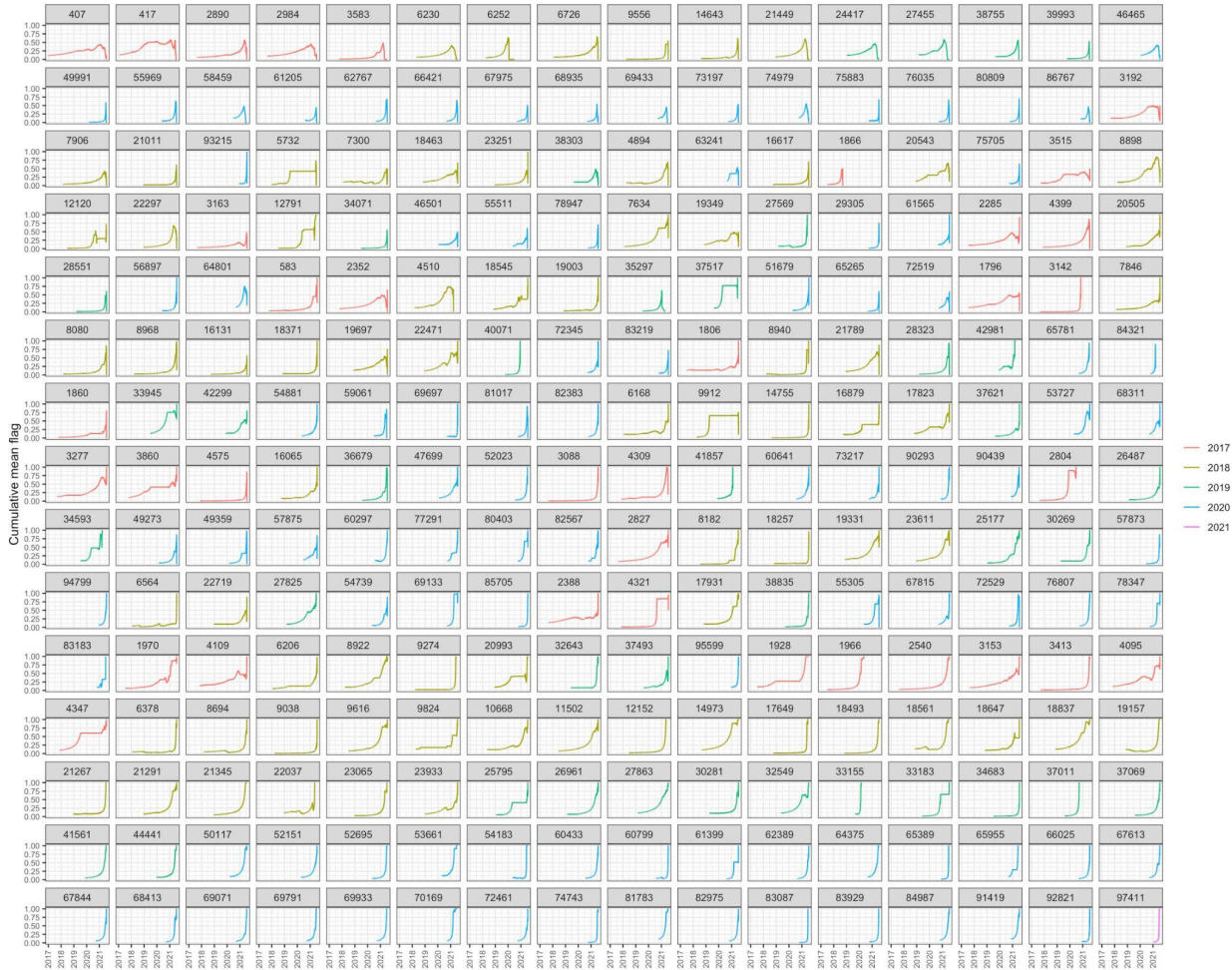

**Figure S4**: *Cumulative mean of the flagged indicator at every hour of measurement calculated by averaging the flag indicator at the hour of measurement and for all subsequent hours for permanently degraded monitors (defined as sensors that record a cumulative mean flagged indicator > 0.4 for at least 100 hours of operation). The title of each subplot corresponds to the PurpleAir monitor ID*

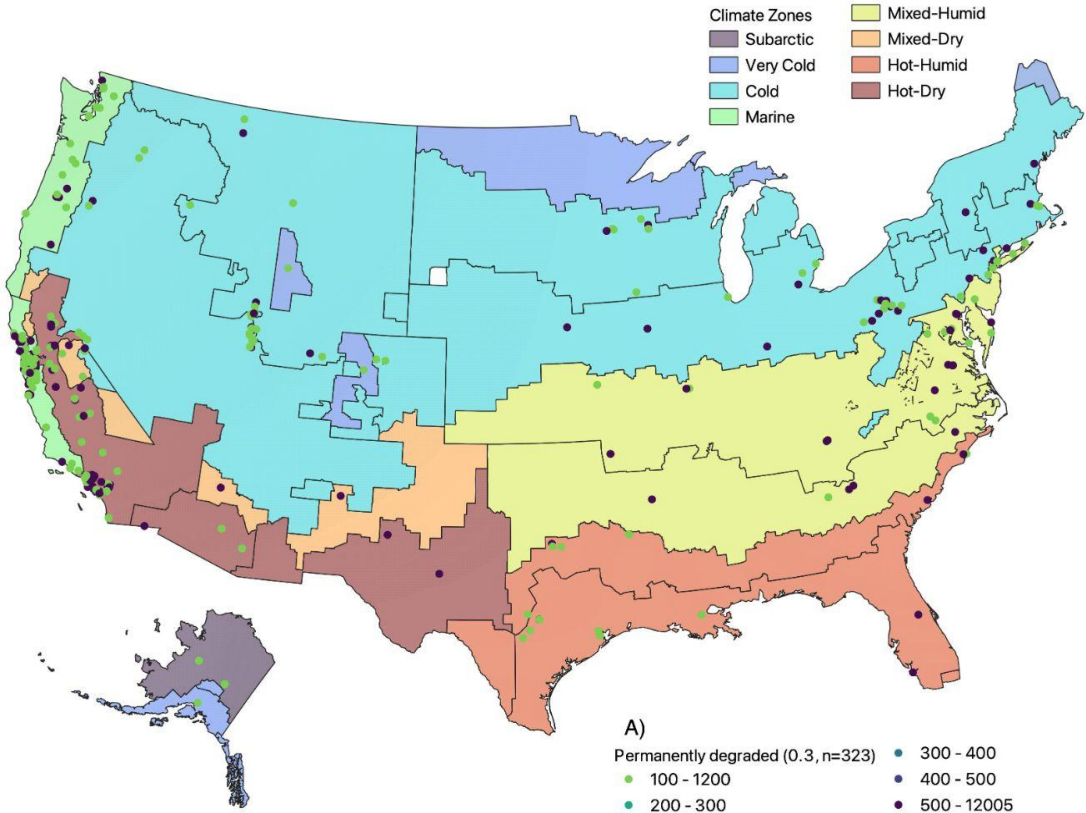

A)

**Climate Zones**
Subarctic · Very Cold · Cold · Marine · Mixed-Humid · Mixed-Dry · Hot-Humid · Hot-Dry

Permanently degraded (0.3, n=323)
- 100 - 1200
- 200 - 300
- 300 - 400
- 400 - 500
- 500 - 12005

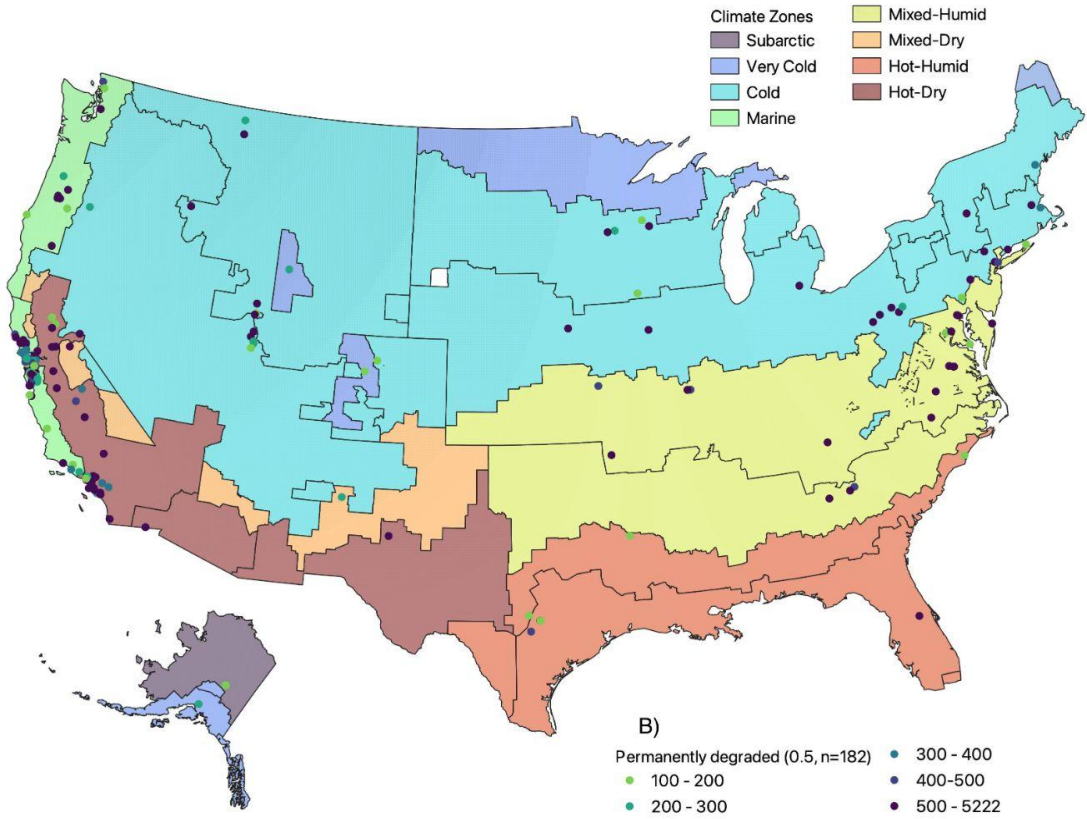

B)

**Climate Zones**
Subarctic · Very Cold · Cold · Marine · Mixed-Humid · Mixed-Dry · Hot-Humid · Hot-Dry

Permanently degraded (0.5, n=182)
- 100 - 200
- 200 - 300
- 300 - 400
- 400-500
- 500 - 5222

**Figure S5**: *Map of permanently degraded PurpleAir monitors with at least 100 measurements for which the cumulative mean of the flag indicator is A)  ≥ 0.3, B) ≥ 0.5. The number of hours of operation for which the cumulative mean of the flag indicator is above the threshold is displayed for each monitor*

# Tables

***Table S1****: Summary statistics of measurements and period of operation of PurpleAir monitors considered in this study by state*

| State | Number of monitors | Number of measurements | Hours of operation per monitor (Min/Max, Mean, Median) |
|---|---|---|---|
| **Alabama** | 18 | 253,307 | Min/Max: 0/36,582 Mean: 15,780 Median: 15,228 |
| **Alaska** | 89 | 1,237,656 | Min/Max:0/30,515 Mean: 11,041 Median:10,585 |
| **Arizona** | 100 | 809,779 | Min/Max:0/34,237 Mean:10,252 Median:7,753 |
| **Arkansas** | 12 | 179,239 | Min/Max:0/28,202 Mean:10,995 Median:10,436 |
| **California** | 7303 | 66,788,259 | Min/Max:0/40,077 Mean:9,798 Median:6,998 |
| **Colorado** | 441 | 3,966,682 | Min/Max:0/34,946 Mean:10,971 Median: 8,847 |
| **Connecticut** | 17 | 194,088 | Min/Max:0/33,610 Mean:12,509 Median:11,306 |
| **Delaware** | 26 | 124,833 | Min/Max:0/34,052 Mean:12,043 Median:13,322 |
| **D.C** | 10 | 90,458 | Min/Max: 0/26,484 Mean: 8,887 Median: 6,861 |
| **Florida** | 49 | 525,499 | Min/Max:0/34,237 Mean: 12,468 Median:11,309 |
| **Georgia** | 60 | 540,393 | Min/Max:0/34,453 Mean: 9,284 Median: 6,912 |
| **Hawaii** | 41 | 637,328 | Min/Max:0/35,119<br>Mean: 13,687<br>Median:13,061 |
| **Idaho** | 75 | 599,536 | Min/Max:0/33,223<br>Mean: 11,968<br>Median:10,336 |
| **Illinois** | 63 | 586,783 | Min/Max:0/34,453<br>Mean: 11,787<br>Median:9,886 |
| **Indiana** | 40 | 591,389 | Min/Max:0/34,786<br>Mean:12,642<br>Median:12,067 |
| **Iowa** | 34 | 393,034 | Min/Max:0/39,633<br>Mean:14,215<br>Median:13,721 |
| **Kansas** | 16 | 586,783 | Min/Max:0/30,126<br>Mean:9,349<br>Median:6,815 |
| **Kentucky** | 8 | 591,389 | Min/Max:0/34,089<br>Mean:12,514<br>Median:10,747 |
| **Louisiana** | 35 | 450,334 | Min/Max:0/31,522<br>Mean:11,766<br>Median:11,218 |
| **Maine** | 33 | 203,152 | Min/Max:0/31,871<br>Mean:9,158<br>Median:7,361 |
| **Maryland** | 143 | 860,122 | Min/Max:0/35,097<br>Mean:11,237<br>Median:8,856 |
| **Massachusetts** | 123 | 775,583 | Min/Max:0/34,688<br>Mean:10,350<br>Median:7,988 |
| **Michigan** | 71 | 637,766 | Min/Max:0/37,905<br>Mean:12,011<br>Median:10,986 |
| **Minnesota** | 66 | 749,580 | Min/Max:0/34,452<br>Mean:12,324<br>Median:11,255 |
| **Mississippi** | 3 | 21,728 | Min/Max:0/21,322<br>Mean:9,763<br>Median:9,183 |
| **Missouri** | 14 | 148,132 | Min/Max:0/32,188<br>Mean: 10,075<br>Median:8,072 |
| **Montana** | 53 | 358,730 | Min/Max:0/34,788<br>Mean:10,902<br>Median:8,160 |
| **Nebraska** | 8 | 131,600 | Min/Max:1/34,592<br>Mean:13,739<br>Median:11,910 |
| **Nevada** | 124 | 1,042,406 | Min/Max:0/28,064<br>Mean:9,045<br>Median:6,670 |
| **New Hampshire** | 18 | 220,990 | Min/Max:0/37,804<br>Mean:15,690<br>Median:15,460 |
| **New Jersey** | 34 | 376,849 | Min/Max:0/35,119<br>Mean:10,891<br>Median:9,235 |
| **New Mexico** | 53 | 594,760 | Min/Max:0/34,669<br>Mean:12,485<br>Median:10,494 |
| **New York** | 113 | 1,026,533 | Min/Max:0/34,090<br>Mean:11,097<br>Median:9,699 |
| **North Carolina** | 176 | 2,105,292 | Min/Max:0/39,123<br>Mean:12,444<br>Median:10,699 |
| **North Dakota** | 4 | 9,824 | Min/Max:1/10,742<br>Mean: 5,927<br>Median:5,831 |
| **Ohio** | 63 | 762,122 | Min/Max:0/34,668<br>Mean:11,430<br>Median: 9,710 |
| **Oklahoma** | 16 | 198,650 | Min/Max:0/35,097<br>Mean:12,808<br>Median:10,674 |
| **Oregon** | 490 | 4,731,279 | Min/Max:0/35,245<br>Mean:10,924<br>Median:8,893 |
| **Pennyslvania** | 242 | 2,700,209 | Min/Max:0/34,450<br>Mean:12,364<br>Median:11,210 |

| Rhode Island | 7 | 97,820 | Min/Max:1/24,700<br>Mean: 12,533<br>Median:12,378 |
| --- | --- | --- | --- |
| **South Carolina** | 19 | 249,528 | Min/Max:0/33,611<br>Mean:11,501<br>Median:9,881 |
| **South Dakota** | 5 | 123,915 | Min/Max:0/34,687<br>Mean:16,455<br>Median:16,431 |
| **Tennessee** | 38 | 378,017 | Min/Max:0/32,475<br>Mean:10,606<br>Median:8,541 |
| **Texas** | 215 | 2,216,316 | Min/Max:0/34,786<br>Mean:11,973<br>Median:10,260 |
| **Utah** | 492 | 7,640,288 | Min/Max:0/39,873<br>Mean:13,670<br>Median:12,735 |
| **Vermont** | 10 | 121,301 | Min/Max:0/34,687<br>Mean:12,310<br>Median:9,612 |
| **Virginia** | 80 | 910,758 | Min/Max:0/35,097<br>Mean:11,954<br>Median:10,989 |
| **Washington** | 677 | 5,787,583 | Min/Max:0/35,244<br>Mean:10,324<br>Median: 7,676 |
| **West Virgina** | 16 | 183,751 | Min/Max:0/34,689<br>Mean:12,177<br>Median:10,178 |
| **Wisconsin** | 56 | 604,028 | Min/Max:0/33,275<br>Mean:11,427<br>Median: 10,189 |
| **Wyoming** | 33 | 156,177 | Min/Max:0/32,212<br>Mean: 6,101<br>Median: 4,678 |

**Table S2**: Summary statistics of measurements and period of operation of PurpleAir monitors considered in this study by climate

| | All | Co-located |
| --- | --- | --- |

| | Number of monitors | Number of measurements | Hours of operation per monitor (Min/Max, Mean, Median) | Number of monitors | Number of measurements | Hours of operation per monitor (Min/Max, Mean, Median) |
|---|---|---|---|---|---|---|
| **Cold** | 2,458 | 26,704,292 | Min/Max:0/39,873<br>Mean: 12,163<br>Median: 10,795 | 39 | 337,026 | Min/Max:0/37,803<br>Mean: 14,023<br>Median: 13,268 |
| **Hot-Dry** | 2,680 | 27,210,216 | Min/Max:0/40,077<br>Mean: 11,487<br>Median:9,404 | 72 | 590,673 | Min/Max:0/35,289<br>Mean:15,451<br>Median: 14,439 |
| **Hot-Humid** | 281 | 3,243,029 | Min/Max:0/37,281<br>Mean: 12,273<br>Median: 11,027 | 1 | 15,432 | Min/Max:0/21,661<br>Mean: 12,893<br>Median: 12,572 |
| **Marine** | 4,842 | 41,448,103 | Min/Max: 0/39,892<br>Mean: 8,644<br>Median: 6,178 | 21 | 305,818 | Min/Max: 0/34,594<br>Mean: 14,644<br>Median: 13,599 |
| **Mixed-Dry** | 361 | 3,527,237 | Min/Max: 0/37,805<br>Mean: 12,197<br>Median: 10,786 | 2 | 34,995 | Min/Max:1/26,359<br>Mean: 15,424<br>Median: 16,110 |
| **Mixed-Humid** | 750 | 6,959,785 | Min/Max: 0/39,123<br>Mean: 11,389<br>Median: 9,434 | 16 | 143,334 | Min/Max: 0/30,827<br>Mean: 9,984<br>Median: 8,448 |
| **Subarctic** | 58 | 839,563 | Min/Max:0/30,029<br>Mean:11,324<br>Median: 11,054 | - | - | - |
| **Very Cold** | 108 | 991,326 | Min/Max:0/34,946<br>Mean: 11,242<br>Median: 10,326 | - | - | - |
| **NA** | 394 | 3,336,389 | Min/Max:0/36,294<br>Mean: 9,684<br>Median: 6,709 | - | - | - |
| **Outside** | | | | | | |
| | Number of monitors | Number of measurements | Hours of operation per monitor (Min/Max, Mean, Median) | - | - | - |
| **Cold** | 2,333 | 25,584,998 | Min/Max:0/39,873<br>Mean: 12,165<br>Median: 10,790 | - | - | - |
| **Hot-Dry** | 2,464 | 25,154,043 | Min/Max: 0/40,077<br>Mean: 11,547<br>Median: 9,473 | - | - | - |
| **Hot-Humid** | 268 | 3,134,869 | Min/Max: 0/37,281<br>Mean: 12,407<br>Median: 11,182 | - | - | - |
| **Marine** | 4,372 | 37,247,271 | Min/Max:0/39,892<br>Mean: 8,627<br>Median: 6,155 | - | - | - |
| **Mixed-Dry** | 337 | 3,325,742 | Min/Max: 0/37,805<br>Mean: 12,064<br>Median: 10,584 | - | - | - |
| **Mixed-Humid** | 697 | 6,590,197 | Min/Max: 0/39,123<br>Mean: 11,539<br>Median: 9,630 | - | - | - |
| **Subarctic** | 57 | 835,664 | Min/Max: 0/30,029<br>Mean: 11,347<br>Median: 11,096 | - | - | - |
| **Very Cold** | 106 | 968,988 | Min/Max: 0/34,946<br>Mean: 11,295<br>Median: 10,383 | - | - | - |
| **NA** | 363 | 3,106,013 | Min/Max:0/36,294<br>Mean: 9,766<br>Median: 6,810 | - | - | - |

**Table S3**: *Performance of the correction models as captured using root mean square error (RMSE), and Pearson correlation (R ). LOSO CV was used to prevent overfitting in the random forest models.* A detailed description of the correction models proposed can be found in (deSouza et al., 2022). Model 2 corresponds to the correction model described using Equation (1)

| Model Number | Name | Equation | Evaluating Correction Model | | Summary of error:<br>Mean (Median) (µg/m³) | |
|---|---|---|---|---|---|---|
| | | | **R** | **RMSE (µg/m³)** | | |
| | **Raw PurpleAir measurements** | | | | | |

| 0 | Raw | | 0.88 | 12.5 | 3.89 (0.95) | 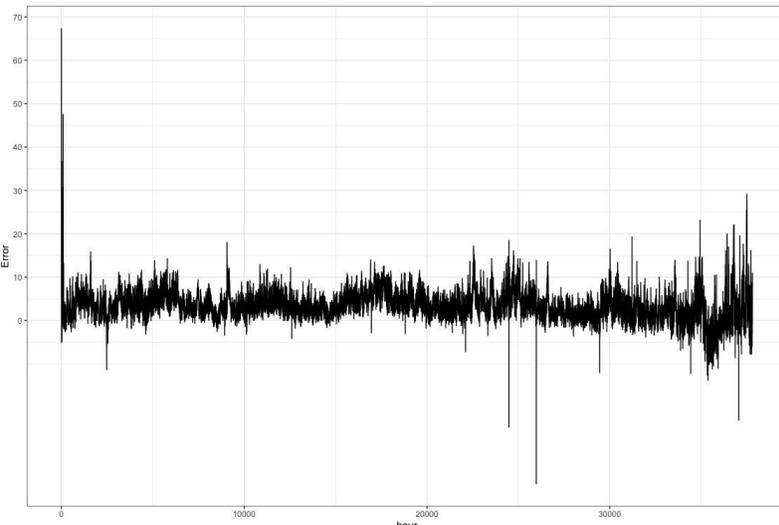 |
| --- | --- | --- | --- | --- | --- | --- |
| | **Multivariate Regression (LOSO CV)** | | | | | |
| 1 | Linear | $PM_{2.5, corrected} = PM_{2.5} \times s1 + b + \varepsilon$ | 0.88 | 6.8 | 3.4 (2.2) | 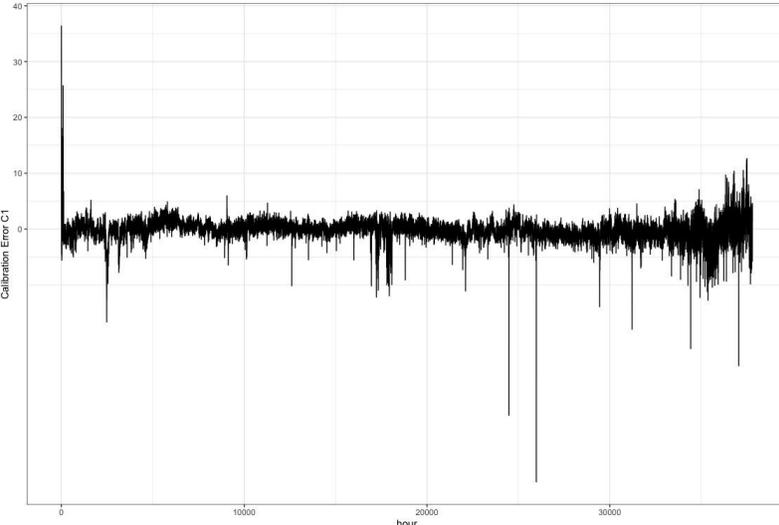 |

| 2 | +RH | $PM_{2.5,\ corrected} = PM_{2.5} \times s_1 + RH \times s_2 + b\ +\ \varepsilon$ | 0.89 | 6.6 | 3.3 (2.2) | 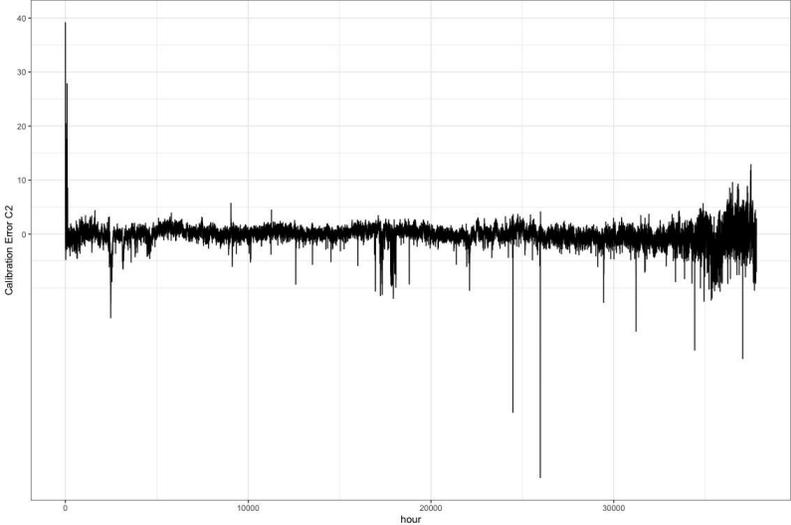 |
| 3 | +T | $PM_{2.5,\ corrected} = PM_{2.5} \times s_1 + T \times s_2 + b\ +\ \varepsilon$ | 0.88 | 6.7 | 3.3 (2.2) | 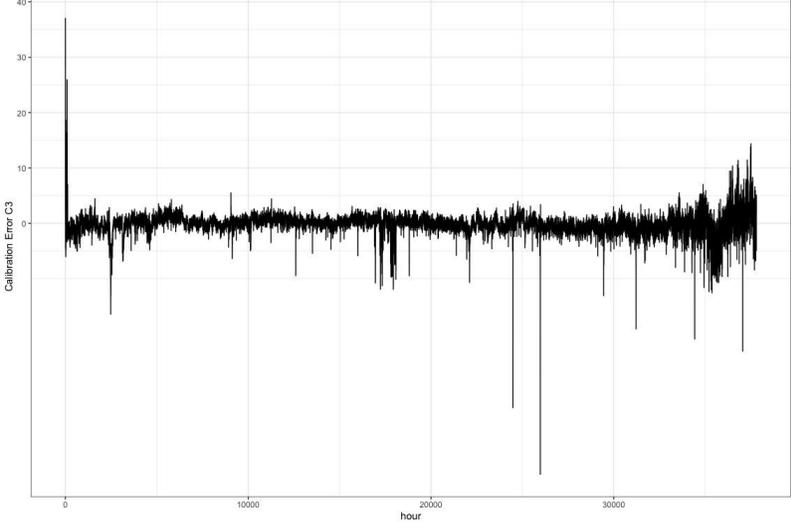 |

| 4 | +RH x T | $PM_{2.5, corrected} = PM_{2.5} \times s_1 + RH \times s_2 + T \times s_3 + RH \times T \times s_4 + b + \varepsilon$ | 0.89 | 6.6 | 3.3 (2.3) |  |
|---|---------|--------|------|-----|-----------|---------|
| 5 | PM x RH | $PM_{2.5, corrected} = PM_{2.5} \times s_1 + RH \times s_2 + RH \times PM_{2.5} \times s_3 + b + \varepsilon$ | 0.89 | 6.4 | 3.1 (2.0) |  |

| 6 | PM x T | $PM_{2.5, corrected} = PM_{2.5}$ x $s_1 + T$ x $s_2 + T$ x $PM_{2.5}$ x $s_3 + b + \varepsilon$ | 0.88 | 6.7 | 3.3 (2.2) |  |
|---|---|---|---|---|---|---|
| 7 | PM x nonlinear RH | $PM_{2.5, corrected} = PM_{2.5}$ x $s_1 + RH^2/(1-RH)$ x $s_2 + RH^2/(1-RH)$ x $PM_{2.5}$ x $s_3 + b + \varepsilon$ | 0.88 | 6.7 | 3.3 (2.2) |  |

| 8 | PM x RH x T | $PM_{2.5,\ corrected} = PM_{2.5} \times s_1 + RH \times s_2 + T \times s_3 + PM_{2.5} \times RH \times s_4 + PM_{2.5} \times T \times s_5 + RH \times T \times s_6 + PM_{2.5} \times RH \times T \times s_7 + b + \varepsilon$ | 0.90 | 6.2 | 3.1 (2.1) | 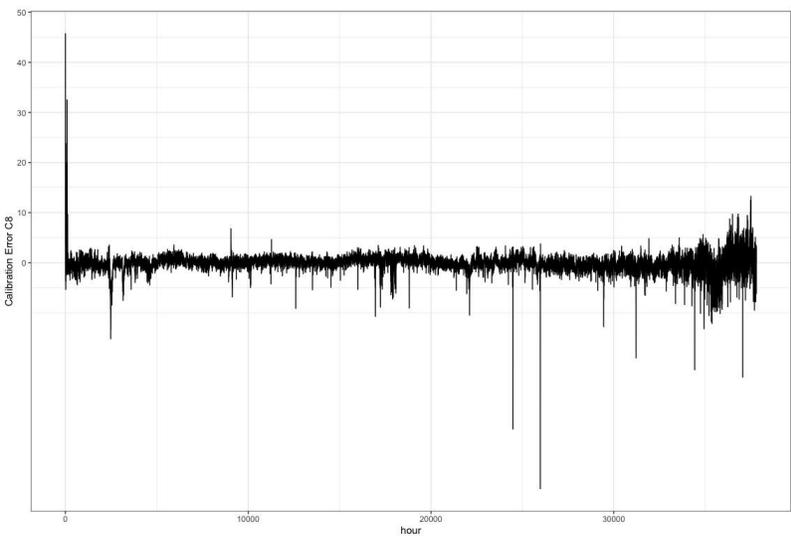 |

**Machine Learning (LOSO CV)**

| 9 | Random Forest | $PM_{2.5,\ corrected} = f(PM_{2.5}, T, RH)$ | 0.99 | 2.4 | 1.3 (0.9) | 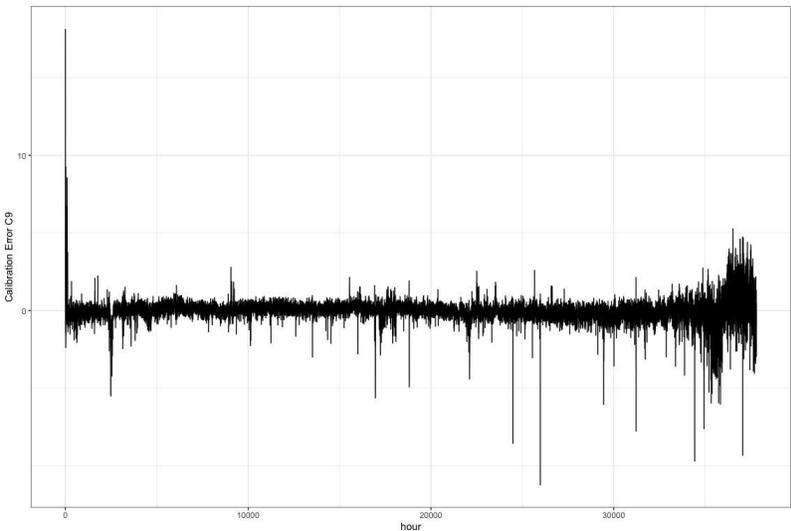 |

***Table S4***: Associations between the correction error and year of operation

| Dataset | Associations (95% Confidence Interval) | | | | | | | | |
|---|---|---|---|---|---|---|---|---|---|
| | Model 1 | Model 2 | Model 3 | Model 4 | Model 5 | Model 6 | Model 7 | Model 8 | Model 9 |
| **All** | -0.28* (-0.29, -0.27) | -0.12* (-0.13, -0.11) | -0.17* (-0.18, -0.16) | -0.08* (-0.10, -0.07) | -0.07* (-0.08, -0.06) | -0.16* (-0.17, -0.15) | -0.23* (-0.24, -0.22) | -0.04* (-0.05, -0.03) | -0.05* (-0.05, -0.05) |
| **Climate Zone (Outside Devices Only)** | | | | | | | | | |
| **Cold** | -0.27* (-0.29, -0.24) | -0.27* (-0.29, -0.25) | -0.19* (-0.21, -0.17) | -0.22* (-0.24, -0.19) | -0.205* (-0.224, -0.185) | -0.20* (-0.22, -0.18) | -0.22* (-0.24, -0.20) | -0.14* (-0.16, -0.12) | -0.08* (-0.09, -0.07) |
| **Hot-Dry** | 0.04* (0.02, 0.05) | 0.08* (0.06, 0.09) | 0.09* (0.08, 0.11) | 0.10* (0.09, 0.12) | 0.13* (0.11, 0.14) | 0.11* (0.09, 0.13) | 0.04* (0.03, 0.06) | 0.13* (0.11, 0.14) | 0.02 (-0.00, 0.01) |
| **Hot-Humid** | -0.86* (-1.04, -0.68) | -0.92* (-1.10, -0.75) | -0.89* (-1.06, -0.71) | -0.93* (-1.10, -0.76) | -0.88* (-1.05, -0.71) | -0.89* (-1.06, -0.71) | -0.86* (-1.04, -0.69) | -0.90* (-1.06, -0.73) | -0.35* (-0.41, -0.28) |
| **Marine** | -0.40* (-0.43, -0.38) | -0.13* (-0.15, -0.10) | -0.27* (-0.29, -0.24) | -0.07* (-0.09, -0.04) | -0.07* (-0.10, -0.05) | -0.24* (-0.27, -0.22) | -0.29* (-0.32, -0.27) | -0.01 (-0.04, 0.01) | -0.02* (-0.03, -0.01) |
| **Mixed-Dry** | -0.93* (-1.03, -0.84) | -0.31* (-0.40, -0.21) | -0.53* (-0.63, -0.44) | -0.28* (-0.37, -0.18) | -0.27* (-0.36, -0.18) | -0.56* (-0.66, -0.47) | -0.83* (-0.92, -0.74) | -0.29* (-0.37, -0.20) | -0.34* (-0.37, -0.30) |
| **Mixed-Humid** | -0.46* (-0.51, -0.41) | -0.28* (-0.33, -0.23) | -0.32* (-0.37, -0.27) | -0.20* (-0.25, -0.16) | -0.20* (-0.25, -0.16) | -0.29* (-0.34, -0.24) | -0.39* (-0.44, -0.35) | -0.13* (-0.17, -0.08) | -0.01* (-0.03, -0.00) |

(*$p < 0.05$)

**Table S5**: *Results from regressing hour and interaction of hour and the cumulative number of high $PM_{2.5}$ measurements recorded on correction error*

| | Coefficients (95% CI) | | |
|---|---|---|---|
| **Intercept** | -1566* (-1774, -1357) | -382.2* (-609.7, -154.7) | 801.9* (618.5, 985.3) |
| **Hour** | -0.18* (-0.19, -0.17) | -0.39* (-0.40, -0.37) | -0.05* (-0.06, -0.04) |
| **Cumulative $PM_{2.5} > 50$** | 29.5* (28.8, 30.1) | - | - |
| **Cumulative $PM_{2.5} > 100$** | - | 2.2* (2.2, 2.3) | - |
| **Cumulative $PM_{2.5} > 500$** | - | - | 864.4* (832.0, 896.7) |
| **Hour: Cumulative $PM_{2.5} > 50$** | $-9.0 \times 10^{-4}$* ($-9.2 \times 10^{-4}$, $-8.7 \times 10^{-4}$) | - | - |
| **Hour: Cumulative $PM_{2.5} > 100$** | - | $-4.9 \times 10^{-5}$* ($-5.2 \times 10^{-5}$, $-4.7 \times 10^{-5}$) | - |
| **Hour: Cumulative $PM_{2.5} > 500$** | - | - | -0.06* (-0.07, -0.06) |

(*$p < 0.05$)

**Table S6**: *Results from regressing hour and interaction of hour and the cumulative number of high $PM_{2.5}$ measurements recorded on normalized correction error*

| | Coefficients (95% CI) | | |
|---|---|---|---|
| **Intercept** | 7,306* (2,125, 12,488) | 6,601* (1,842, 11,359) | 8,263* (3,727, 12,799) |
| **Hour** | 0.6* (0.3, 0.9) | 0.5* (0.2, 0.8) | 0.5* (0.2, 0.8) |
| **Cumulative PM$_{2.5}$ > 50** | 2.9<br>(-14.1, 19.8) | - | - |
| **Cumulative PM$_{2.5}$ > 100** | - | 90<br>(19, 160) | - |
| **Cumulative PM$_{2.5}$ > 500** | - | - | -238.9<br>(-1039.4, 561.6) |
| **Hour: Cumulative PM$_{2.5}$ > 50** | -1.6 x 10$^{-4}$<br>(-8.3 x 10$^{-4}$, 5.1 x 10$^{-4}$) | - | - |
| **Hour: Cumulative PM$_{2.5}$ > 100** | | -2.6 x 10$^{-3}$<br>(-5.5 x 10$^{-3}$, 0.2 x 10$^{-3}$) | - |
| **Hour: Cumulative PM$_{2.5}$ > 500** | - | - | 8.2 x 10$^{-3}$<br>(-5.0 x 10$^{-2}$, 6.7 x10$^{-2}$) |

(*$p < 0.05$)